\documentclass[aps, twocolumn, superscriptaddress, showpacs, nofootinbib, longbibliography]{revtex4-1}

\usepackage[utf8]{inputenc}
\usepackage[T1]{fontenc}
\usepackage{ae,aecompl} 
\usepackage{graphicx}
\graphicspath{{figures/}} % Directory in which figures are stored
\usepackage{makecell}
\usepackage{amsmath}
\usepackage{xcolor}
\usepackage{amssymb}
\usepackage{latexsym}
\usepackage{wasysym}
\usepackage{psfrag}
\usepackage{ifthen}
\usepackage{hyperref}
\hypersetup{citecolor=blue,colorlinks=true}
\usepackage{longtable}
\usepackage{float}
\usepackage[utf8]{inputenc}
\usepackage{lineno}
\usepackage{units}
\usepackage[title]{appendix}
\usepackage{aas_macros}
\usepackage{ulem}
\usepackage{subfigure}
\usepackage{multirow}
\usepackage{diagbox}
\def\mnras{Mon.~Not.~Roy.~Astron.~Soc.}
\def\apjl{Astrophys.~J.~Lett.}
\def\apj{Astrophys.~J.}
\def\prd{Phys.~Rev.~D}
\def\aap{A~\&~A}

\definecolor{maob}{rgb}{0.6,0.1,0.2}

\newcommand{\msun}{$\mathrm{M_{\odot}}$}

\newcommand{\vecrk}{\left .\vec{r}\right .^k}

\begin{document}
%\linenumbers

\title{Inference of proto-neutron star properties in core-collapse supernovae from a gravitational-wave detector network}

\author{Tristan~Bruel}%\orcidlink{0000-0002-1789-7876}}
\affiliation{Universit\'e C\^ote d'Azur, Observatoire C\^ote d'Azur, CNRS, Artemis, F-06304 Nice, France}
\affiliation{Universit\'e C\^ote d'Azur, Observatoire C\^ote d'Azur, CNRS, Lagrange, F-06304 Nice, France}

\author{Marie-Anne~Bizouard}%\orcidlink{0000-0002-4618-1674}}
\affiliation{Universit\'e C\^ote d'Azur, Observatoire C\^ote d'Azur, CNRS, Artemis, F-06304 Nice, France}

\author{Martin~Obergaulinger}%\orcidlink{0000-0001-5664-1382}}
\affiliation{Departamento de Astronom\'{\i }a y Astrof\'{\i }sica, Universitat de Val\`encia, E-46100 Burjassot, Val\`encia, Spain}

\author{Patricio~Maturana-Russel}%\orcidlink{0000-0002-5211-9818}}
\affiliation{Department of Statistics, The University of Auckland, Auckland, New Zealand}
\affiliation{Department of Mathematical Sciences, Auckland University of Technology, Auckland, New Zealand}

\author{Alejandro~Torres-Forn\'e}%\orcidlink{0000-0001-8709-5118}}
\affiliation{Departamento de Astronom\'{\i }a y Astrof\'{\i }sica, Universitat de Val\`encia, E-46100 Burjassot, Val\`encia, Spain}

\author{Pablo~Cerd\'a-Dur\'an}%\orcidlink{0000-0003-4293-340X}}
\affiliation{Departamento de Astronom\'{\i }a y Astrof\'{\i }sica, Universitat de Val\`encia, E-46100 Burjassot, Val\`encia, Spain}
\affiliation{Observatori Astron\`omic, Universitat de Val\`encia, E-46980, Paterna, Val\`encia, Spain} 

\author{Nelson~Christensen}%\orcidlink{0000-0002-6870-4202}}
\affiliation{Universit\'e C\^ote d'Azur, Observatoire C\^ote d'Azur, CNRS, Artemis, F-06304 Nice, France}
\affiliation{Physics and Astronomy, Carleton College, Northfield, MN 55057, USA}

\author{Jos\'e~A.~Font}%\orcidlink{0000-0001-6650-2634}}
\affiliation{Departamento de Astronom\'{\i }a y Astrof\'{\i }sica, Universitat de Val\`encia, E-46100 Burjassot, Val\`encia, Spain}
\affiliation{Observatori Astron\`omic, Universitat de Val\`encia, E-46980, Paterna, Val\`encia, Spain} 

\author{Renate~Meyer}%\orcidlink{0000-0003-0268-8569}}
\affiliation{Department of Statistics, The University of Auckland, Auckland, New Zealand}

%\today
\begin{abstract}
The next Galactic core-collapse supernova (CCSN) will be a unique opportunity to study within a fully multi-messenger approach the explosion mechanism responsible for the formation of neutron stars and stellar-mass black holes. State-of-the-art numerical simulations of those events reveal the complexity of the gravitational-wave emission which is highly stochastic. This challenges the possibility to infer the properties of the compact remnant and of its progenitor using the information encoded in the waveforms. In this paper we take further steps in a program we recently initiated to overcome those difficulties. In particular we show how oscillation modes of the proto-neutron star (PNS), highly visible in the gravitational-wave signal, can be used to reconstruct the time evolution of their physical properties. Extending our previous work where only the information from a single detector was used we here describe a new data-analysis pipeline that coherently combines gravitational-wave detectors' data and infers the time evolution of a combination of the mass and radius of the compact remnant. The performance of the method is estimated employing waveforms from 2D and 3D CCSN simulations covering a progenitor mass range between 11$\mathrm{M_{\odot}}$\, and 40$\mathrm{M_{\odot}}$\, and different equations of state for both a network of up to five second-generation detectors and the proposed third-generation detectors Einstein Telescope and Cosmic Explorer. Our study shows that it will be possible to infer PNS properties for CCSN events occurring in the vicinity of the Milky Way, up to the Large Magellanic Cloud, with the current generation of gravitational-wave detectors.
\end{abstract}

\maketitle

\section{Introduction}
\label{sec:introduction}
A new astronomy has emerged with the detection of gravitational waves (GWs) from stellar mass black hole and neutron star mergers~\cite{Abbott:2016aoc,Abbott:2017vwq}. GWs are providing unperturbed information about the dynamics of the most energetic events of the Universe associated with compact objects that can be complemented by the observation of electromagnetic counterparts that probe the interaction of the ejected matter with the environment~\cite{Abbott:2017zic,Abbott:2017ync}. Thanks to their continuing sensitivity improvements, the Advanced LIGO~\cite{Aasi:2014jea}, Advanced Virgo~\cite{Acernese:2014hva} and KAGRA~\cite{Aso:2013eba} detectors will explore a constantly increasing volume of the Universe for the next decade~\cite{KAGRA:2013rdx}, before proposed third generation ground-based detectors such as Einstein Telescope~\cite{Punturo:2010zza} or Cosmic Explorer~\cite{reitze2019cosmic}, expected to be $\sim 10$\, times more sensitive, become operational. Moreover, the kHz-band detector NEMO~\cite{Ackley2020} has also recently been proposed in Australia to study specifically nuclear matter using GWs.
In this bright panorama, detecting the GWs emitted by a Galactic core-collapse supernova (CCSN) is one of the main next challenges for GW astronomy \cite{Sedda_2020}. The event is expected to be accompanied with neutrinos and a delayed electromagnetic emission in all bands. The latter, however, could be absent if a black hole forms before the star explodes. Each of these messengers would actually {signal} the occurrence of such a rare event ($2-3$ per century and galaxy ~\cite{Rozwadowska:2021}) and provide precise timing and sky location information~\cite{SNEWS:2020tbu}.
Current GW detectors have the sensitivity to detect and localize a CCSN only if it happens in the Galaxy~\cite{Gossan:2016,Szczepanczyk:2021bka}.

Being such highly dynamical events CCSN have always been prime candidates for GW emission. At the end of the life of most massive stars ($M \gtrsim 9$ \msun\, at zero age main sequence) the core, mainly composed of iron nuclei, has a mass {close to} the Chandrasekhar mass limit. Gravity is no longer compensated by the pressure of  elativistic degenerate electrons and the collapse sets in.
When supra-nuclear density is reached, nuclear matter stiffens and the core bounces generating a shock wave propagating into the still in-falling matter. Iron nuclei are photo-dissociated costing energy to the shock wave that quickly halts at $\sim$ \unit[150]{km} radius.
In most situations a fraction of the energy of the neutrinos emitted at high densities is expected to be deposited in the region behind the shock, reviving it and driving an explosion that unbinds the stellar envelope. The neutrino emission and absorption processes that lead to the explosion can be enhanced by hydrodynamic instabilities in the post-shock region. This neutrino-driven convection is suspected to create funnels of accretion onto the proto-neutron star (PNS) that excite its oscillation modes. See \cite{Vartanyan_2023} for more details on the turbulent accretion onto the PNS and the importance of late-time 3D simulations to capture the complete GW signal. For a small fraction of progenitor stars, the core has a significant rotation at the time of the collapse and, additionally to the neutrino energy deposition, magnetic fields are able to transfer rotational kinetic energy of the PNS to the shock, leading to more energetic explosions. We refer the reader to \cite{Pajkos_2021} for more details on CCSN parameter estimation methods in the case of rotating sources.

The gravitational waveforms extracted from multi-dimensional numerical simulations of CCSN consistently exhibit very complex features~\cite{Abdikamalov2020,Mezzacappa:2022hmk,Radice:2018usf,Kuroda:2016,Cerda:2013,powell:2019,Andresen:2017,mueller:13gw,Murphy:09,Morozova:2018,VartanyanBurrows_2020}. Among them, the presence of a rising, high-frequency feature in the GW spectrogram is especially noticeable. This is associated with an oscillation mode of the PNS and is connected in particular to a region of the PNS surface with positive Brunt-V\"{a}is\"{a}l\"{a} frequency. The nature of this mode remains a matter of discussion. It has been interpreted as either a g-mode~\cite{mueller:13gw,Cerda:2013,Torres:2018,Torres:2019b} or an f-mode~\cite{Morozova:2018,Sotani:2019,Vartanyan_2023}.
The precise characterization of the remnant frequencies is an important task as it is expected to allow for PNS asteroseismology using GW information. This approach aims at providing universal relations between the mode frequencies and the physical parameters of the PNS that barely depend on the progenitor model or the equation of state~\citep{Sotani:2017ubz,Torres:2019b,Sotani:2021ygu}.

So far, most of the inference efforts have focused on determining the nature of the explosion mechanism and rely on catalogues of CCSN waveforms from numerical simulations~\citep[e.g.][]{roever:09,Logue:2012zw,powell:2016,Ainara:2022}. Correlations between GWs, neutrino emissions and some of the CCSN physical properties have been found in \cite{Vartanyan_2019} using 3D simulations. A method to estimate the time evolution of the PNS mass and radius from the CCSN neutrino signal has also been proposed in \cite{Nagaruka_2022}.
A recent study has explored Bayesian inference techniques parameterizing the high-frequency GW signal with asymmetric chirplet waveforms in an attempt to extract the PNS physical parameters~\cite{Powell:2022}. In~\cite{Bizouard:2020sws} we proposed a method to extract the time evolution of a combination of the PNS mass and radius using the universal relations derived in~\cite{Torres:2019b} that does not rely on a specific waveform model. This study, which considered only single-detector data, showed that it is possible to infer PNS properties for a galactic source using Advanced LIGO and Advanced Virgo data at design sensitivities. The current paper is an extension of the study initiated in~\cite{Bizouard:2020sws}. Here we describe a new data-analysis pipeline that coherently combines data from a network of GW detectors to reconstruct PNS properties. We show that employing a network of detectors has a significant impact in the inference prospects, especially when accounting for third-generation detectors. We also note that if the study is restricted to the current generation of GW detectors, the possibilities to perform asteroseismology of PNS are already enhanced when using a network of detectors in comparison to the single-detector case, extending the coverage of the inference to the vicinity of the Milky Way, up to the Large Magellanic Cloud.

The paper is organized as follows. Section~\ref{sec:network} presents a brief description of the ground-based interferometers we consider. In Section~\ref{sec:strategy} we summarize the multi-messenger observational prospects of detecting a CCSN event in the next decades. The numerical simulations and GW data used in this study are discussed in Section~\ref{sec:simulations}. Section~\ref{sec:method} describes the data analysis pipeline whose performance is evaluated in Section~\ref{sec:results}. Finally, Section~\ref{sec:conclusions} presents our conclusions and outlines possible extensions of this work. 

\section{Network of detectors}
\label{sec:network}

The current generation of advanced GW detectors (Advanced LIGO~\cite{Aasi:2014jea}, Advanced Virgo~\cite{Acernese:2014hva} and KAGRA~\cite{Aso:2013eba}) have been progressively put in operation since 2015. The two LIGO detectors in the USA (LIGO Hanford and LIGO Livingston) have 4-km long arms while the Virgo detector in Italy and the KAGRA detector in Japan have 3-km long arms. This network of detectors will be completed by a third LIGO-like detector in India (LIGO Aundha) around 2027~\cite{Saleem:2021iwi}. Despite their optical and control configuration differences, all these L-shape detectors are fully characterized by their sensitivity expressed in terms of their noise amplitude spectral density (ASD) and by their position and orientation on Earth. They operate as a network with coordinated observing runs in between periods of upgrades and commissioning that aim at bringing the detectors to their design sensitivity circa 2027~\cite{Abbott:2020xxx}.
In addition to this network of existing facilities, a third generation of ground-based detectors are already proposed for the next decade. The two main projects are Einstein Telescope (ET)~\cite{Punturo:2010zza} in Europe and Cosmic Explorer (CE)~\cite{reitze2019cosmic,Evans:2021gyd} in the USA. The current design for ET\footnote{https://www.et-gw.eu} is based on a triangular-shape detector composed of six 10-km long interferometers with an angle of \unit[60]{degrees} between arms, underground and with low-temperature test masses / low laser power and room temperature test masses / high laser power mixed configuration to provide good sensitivity at very low and high frequency. The CE project\footnote{https://cosmicexplorer.org} is currently featuring two sites hosting L-shape interferometers with extra-long arms of respectively 40 and \unit[20]{km}.

The response of any of these interferometers to a GW depends on the direction of propagation of the GW, its polarization, and the position and orientation of the detector's arms. The signal strain amplitude is given by
\begin{equation}\label{eq:response1}
    h(t) = F_{+}(\vec{r}, \theta, \phi, \psi) h_{+}(t) + F_{\times}(\vec{r}, \theta, \phi, \psi) h_{\times}(t)
\end{equation}
where $F_+^k$ and $F_{\times}^k$ are the detector response functions, while $h_{+}(t)$ and $h_{\times}(t)$ are the GW polarizations given in the source propagation frame. (Each detector is labelled by index $k$.) The angles $\theta$ and $\phi$ specify the source location in an Earth-fixed coordinate system, and $\psi$ is the wave polarization angle that is unknown. The vector $\vec{r}$ gives the surface position of the detector with respect to the center of the Earth.

\begin{table*}[ht]
    \centering
    \begin{tabular}{cc}
    \begin{tabular}{c|c|c|c|c|c}
        \hline
        Detector & $\lambda$ [\textdegree] & $\phi$ [\textdegree] & $h$ [m] & $\Psi1$ [\textdegree] & $\Psi2$ [\textdegree] \\
        \hline
        LIGO Hanford (H) & -119.41 & 46.46 & 142.55 & 324.00 & 234.00 \\
        LIGO Livingston (L) & -90.77 & 30.56 & -6.57 & 252.28 & 162.28 \\
        Virgo (V) & 10.50 & 43.63 & 51.88 & 19.43 & 289.43 \\
        KAGRA (K) & 137.31 & 36.41 & 414.18 & 60.40 & 330.40 \\
        LIGO Aundha (A) & 77.03 & 19.61 & 0.0 & 332.38 & 242.38 \\
        \hline
    \end{tabular}
    &
    
    \begin{tabular}{c|c|c|c|c|c}
    \hline
        Detector & $\lambda$ [\textdegree] & $\phi$ [\textdegree] & $h$ [m] & $\Psi1$ [\textdegree] & $\Psi2$ [\textdegree] \\
        \hline
        Einstein Telescope 1 (ET) & 10.50 & 43.63 & 0.0 & 89.95 & 29.96 \\
        Einstein Telescope 2 (ET) & 10.63 & 43.63 & 0.0 & 330.04 & 270.05 \\
        Einstein Telescope 3 (ET) & 10.57 & 43.71 & 0.0 & 210.00 & 150.00 \\
        Cosmic Explorer \unit[40]{km} (CE40) & -112.83 & 43.83 & 0.0 & 180.00 & 90.00 \\
        Cosmic Explorer \unit[20]{km} (CE20) & -106.48 & 33.16 & 0.0 & 240.00 & 150.00 \\
        \hline
    \end{tabular}\\
    \end{tabular}
    \caption{Coordinates of the current (left) and future (right) generation of ground-based interferometers. The longitudes $\lambda$ and latitudes $\phi$ give the locations of the beam-splitters/corner stations, $h$ gives the elevation, in meters, above the reference ellipsoid WGS84, and $\Psi1$ and $\Psi2$ are the two arms orientation angles defined clockwise from the local North. The values for LIGO Hanford, LIGO Livingston, Virgo, KAGRA and LIGO Aundha are extracted from~\cite{lalsuite}. The location and orientation of ET and CE are not yet known. For simplicity we have chosen to locate ET at the location of the Virgo detector with an arbitrary orientation. The two CE sites are set according to~\cite{Borhanian:2020ypi}, where a \unit[40]{km} detector is located in Idaho and one of \unit[20]{km} in New Mexico.}
    \label{tab:detectors}
\end{table*}

The configuration of a network of detectors in the Earth-fixed frame is fully determined by the location of their beam-splitters and the orientation vectors of their arms. In Table~\ref{tab:detectors} we give the coordinates of both, the current generation and the third-generation GW detectors that we consider in this study. As the location and orientation of third-generation detectors are still not decided, we have chosen to locate ET at the Virgo location with arbitrary orientation while the CE sites are set in Idaho (USA) and New Mexico (USA)~\cite{Borhanian:2020ypi}.
We use the algebra described in~\cite{Riles:2010} to calculate the antenna pattern functions for each interferometer for a source with known sky position at a given epoch and with a fixed polarization angle.

\begin{figure}
 \centering
 \includegraphics[width=0.5\textwidth]{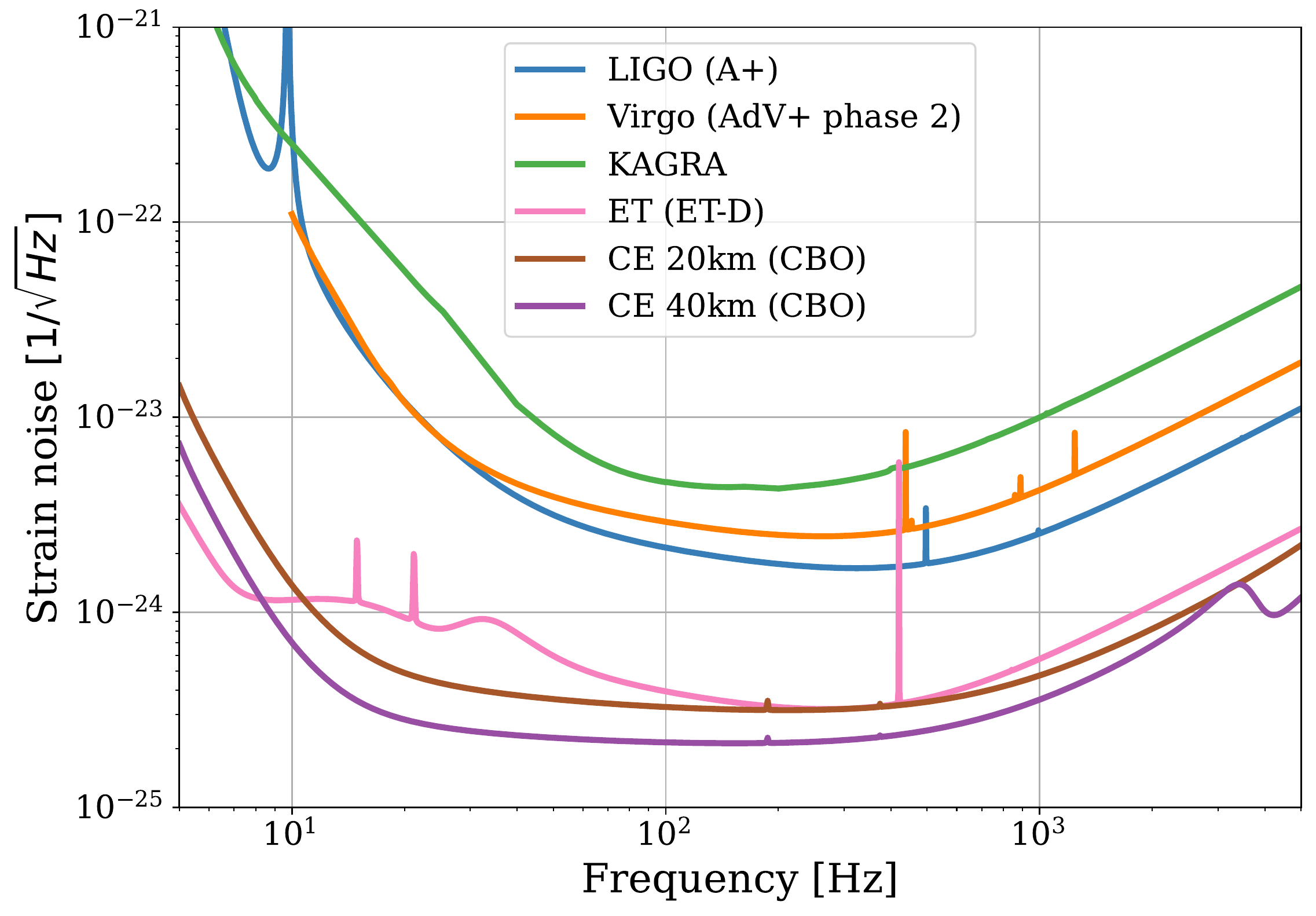}
 \caption{Amplitude spectral densities as a function of frequency for the ground-based interferometers considered in this paper. The LIGO (A+), Virgo (AdV+ Phase 2) and KAGRA ASDs are taken from~\cite{2Gsens_prospects:2020}. Einstein Telescope sensitivity corresponds to the ET-D configuration~\cite{Hild_2011} while the Cosmic Explorer sensitivities correspond to the compact binary optimized configuration~\cite{2022ApJ...931...22S}.} 
 \label{fig:PSDs}
\end{figure}

In addition to their response functions, the detectors are characterised by their noise properties. For the network formed by LIGO Hanford (H), LIGO Livingston (L), LIGO Aundha (A), Virgo (V) and KAGRA (K), we consider the currently projected design sensitivity of the upgraded detectors. Namely for the three LIGO detectors we use the A+ upgrade program sensivity~\cite{aLIGOsens:2018}. For Virgo we use the Advanced Virgo+ Phase 2 sensivity~\cite{Abbott:2020xxx} and for KAGRA we use the design sensitivity~\cite{Abbott:2020xxx}\footnote{https://dcc.ligo.org/LIGO-T2000012/public}. For ET we consider three co-located (triangular) interferometers with the same ET-D configuration ASD~\cite{Hild_2011}\footnote{https://apps.et-gw.eu/tds/ql/?c=12989}, encompassing the cryogenic and room temperature interferometers' sensitivities, while for CE we consider two ASDs corresponding to the two different arm-length detectors. For both CE detectors we use the compact binary optimized sensitivity curves~\cite{2022ApJ...931...22S}\footnote{ https://dcc.cosmicexplorer.org/CE-T2000017/public}. These ASDs, shown in Figure~\ref{fig:PSDs}, are used to generate time-series of Gaussian colored noise for each detector. Injecting a waveform into these time-series allows us to simulate data segments containing a CCSN GW signal buried in realistic noise.

We identify a network of detectors with the letters corresponding to each of its detectors (e.g. HLVK is the network composed of LIGO Hanford (H), LIGO Livingston (L), Virgo (V) and KAGRA (K)).

\section{Observational strategy}
\label{sec:strategy}

CCSNe are such rare events that it is necessary to optimize the GW observation strategy to maximise the detection probability. Both the availability of the GW detectors network and the multi-messenger properties of these astronomical phenomena must be taken into account. Many large field of view robotic surveys, such as DLT40~\cite{Tartaglia:2018}, ASAS-SN~\cite{Shappee:2014}, ZTF~\cite{Bellm:2018}, and Pan-STARRS~\cite{Chambers:2016}, are detecting supernovae in their early explosion phase, and some of them are identified as CCSNe. Although the delayed shock breakout and subsequent light curve measurements do not allow for the estimation of the time of the bounce with an accuracy better than a few days~\cite{Abbott:2019ryq,Gill:2022amm, Barker:2021iyr}, these facilities provide the most accurate (few arc minutes) pointing information when the shock breakout starts emitting UV and X-ray signals.

Depending on their specific characteristics, GW detectors and low-energy neutrino detectors can observe CCSNe in the vicinity of the Galaxy. Many low-energy neutrino detectors (Super-Kamiokande~\cite{Fukuda:2003}, IceCube~\cite{Abbasi:2009}, Km3Net~\cite{Adrian-Martinez:2016}, LVD~\cite{Aglietta:1992}, Borexino~\cite{Alimonti:2009}, KamLAND~\cite{Gando:2016}, JUNO~\cite{An:2016}, SNO+~\cite{Chen:2005}, Baksan~\cite{Kuzminov:2012}) have the capability to detect the neutrino burst emission up to $\sim$ \unit[100]{kpc}~\cite{SNEWS:2020tbu}. While this horizon will be further increased when Hyper-K~\cite{Hyper-Kamiokande:2018} and DUNE~\cite{DUNE:2020} are in operation, it is likely that typical neutrino-driven or magneto-hydrodynamical supernova explosions in Andromeda, the most luminous galaxy of the Local Group, remain out of reach~\cite{Gossan:2016}. Therefore, we have to consider all galaxies nearer than Andromeda, even though most of them are low-luminosity dwarf galaxies containing old stars. The explosion of SN 1987A in one of the satellite galaxies of the Milky Way, the Large Magellanic Cloud (LMC) at a distance of $\sim$ \unit[50]{kpc}, underlines the benefit of increasing the range of the detectors beyond the Galaxy itself, as, despite their limited capabilities at that time, neutrino detectors were able to confirm that $\sim$ 99\% of the energy of the collapse was radiated away by neutrinos~\cite{Hirata:1988ad,Bionta:1987qt}.

Neutrino detectors can localize the source in the sky with a precision that depends on the distance to the source. Crucially for the attempt to extract a possible GW signal from the noise, they will provide the most accurate estimate of the core bounce timestamp with a precision of $\sim$ \unit[10]{ms}~\cite{Pagliaroli:2009,Halzen:2009} which can be further refined to few ms assuming a neutrino flux model. In the case of failed supernova or black hole formation, the water Cherenkov neutrino detector Super-Kamiokande has a pointing accuracy of a few degrees for a source at \unit[10]{kpc}~\cite{Abe:2016kji}. In this study we assume that information from neutrino and electromagnetic signals will be accurate enough to consider that the source sky position and the arrival time of the signal are known without error.

\section{CCSN simulations}
\label{sec:simulations}

\begin{table}[t]
    \centering
    \begin{tabular}{c|ccc|ccc}
  \hline
  Model & $M_\mathrm{ZAMS}$ & progenitor& EOS & $t_{\mathrm{f}}$& $t_{\rm explosion}$ & $M_{\mathrm{PNS, f}}$\\
  name& [\msun] & model & & [$\mathrm{s}$] & [$\mathrm{s}$] & [\msun] 
  \\ 
  \hline
  \texttt{s11} & 11.2 & \cite{Woosley_Heger_Weaver__2002__ReviewsofModernPhysics__The_evolution_and_explosion_of_massive_stars}& LS220 & 1.86 & $\times$ & 1.47 
  \\ 
  \texttt{s15} & 15.0 & \cite{Woosley_Heger_Weaver__2002__ReviewsofModernPhysics__The_evolution_and_explosion_of_massive_stars}& LS220 & 1.66 & $\times$ & 2.00 
    \\ 
  \texttt{s15S} & 15.0 & \cite{Woosley_Heger_Weaver__2002__ReviewsofModernPhysics__The_evolution_and_explosion_of_massive_stars}& SFHo & 1.75 & $\times$ & 2.02 
    \\ 
  \texttt{s15G} & 15.0 &    \cite{Woosley_Heger_Weaver__2002__ReviewsofModernPhysics__The_evolution_and_explosion_of_massive_stars}& GShen & 0.97 & $\times$ & 1.86
     \\ 
  \texttt{s20} & 20.0 & \cite{Woosley_Heger_Weaver__2002__ReviewsofModernPhysics__The_evolution_and_explosion_of_massive_stars}& LS220 & 1.53 & $\times$ & 1.75 
    \\ 
  \texttt{s20S} & 20.0 & \cite{Woosley_Heger__2007__physrep__Nucleosynthesisandremnantsinmassivestarsofsolarmetallicity} & SFHo & 0.87 & $\times$ & 2.05 
  \\ 
  \texttt{s25} & 25.0 & \cite{Woosley_Heger_Weaver__2002__ReviewsofModernPhysics__The_evolution_and_explosion_of_massive_stars}& LS220 & 1.60 & $0.91$ & 2.33 
    \\ 
  \texttt{s40} & 40.0 & \cite{Woosley_Heger_Weaver__2002__ReviewsofModernPhysics__The_evolution_and_explosion_of_massive_stars}& LS220 & 1.70 & $1.52$ & 2.23 
    \\
    \texttt{s15--3De} & 15.0 & \cite{Woosley_Heger_Weaver__2002__ReviewsofModernPhysics__The_evolution_and_explosion_of_massive_stars} & SFHo & 1.30 &  $\times$ & 1.96
    \\ 
    \texttt{s15--3Dp} & 15.0 & \cite{Woosley_Heger_Weaver__2002__ReviewsofModernPhysics__The_evolution_and_explosion_of_massive_stars} & SFHo & 1.30 & $\times$ & 1.96 
    \\ \hline
    \end{tabular}

    \caption{List of axisymmetric 2D and 3D CCSN simulations used to test the performance of the inference method. The last three columns show the post-bounce time at the end of the simulation, the one at the onset of the explosion (non exploding models marked with $\times$), and the PNS mass at the end of the simulation. Note that the last two lines correspond to the same 3D simulation, but we distinguish between the GW emission in the equatorial plane and along the polar axis (\texttt{s15--3De} and \texttt{s15--3Dp}, respectively).}
    \label{tab:simulations}
\end{table}

To test how well we can infer the PNS physical parameters we simulate, for different configurations of the GW detectors network, CCSN signals coming from sources at fixed sky positions and reaching each GW detectors' location on Earth with a proper time delay.
We consider different CCSN signals extracted from 2D and 3D numerical simulations performed with the \texttt{AENUS-ALCAR} code for spectral neutrino-hydrodynamics~\cite{Just:2015}. The first set consists of the eight waveforms of axisymmetric 2D simulations used in \cite{Bizouard:2020sws}. We extract from the simulations both the GW amplitude and the time evolution of the PNS mass, $M_{\rm PNS}$, and its radius, $R_{\rm PNS}$. The progenitor masses range between \unit[11.2]{\msun} and \unit[40]{\msun}, and three different equations of state (EOS) have been considered.
We also employ a second set comprising two waveforms obtained from the 3D simulation of a \unit[15]{\msun} progenitor star, performed with the same code. For this model the core fails to produce an explosion within the \unit[1.3]{s} of simulation time. Continuous accretion onto the PNS causes its mass to grow to a final value of \unit[1.96]{\msun}. None of the multidimensional models includes rotation or magnetic fields. The list of CCSN waveforms used in this paper is given in Table \ref{tab:simulations}.
It is important to underline that the GW amplitudes obtained with the 2D simulations are systematically higher than those of the 3D simulation. This is illustrated in the second column of Table \ref{tab:perf}, which gives the gravitational energy for all the simulations used in this work.

\section{Description of the analysis method}
\label{sec:method}

We generalize the method proposed in~\cite{Bizouard:2020sws} to $N$ GW detectors to fully exploit the sky coverage of the current network of detectors. Following inverse problem and optimal detection methods proposed, for instance, in~\cite{Guersel:1989th,Flanagan:1997kp,Jaranowski:1998qm,Pai:2000zt,Anderson:2000yy,Klimenko:2005xv,Sutton:2009gi,Harry:2010fr,Klimenko:2015ypf}, we coherently combine in a likelihood ratio function the data initially transformed in the time-frequency (TF) domain. The exercise is greatly simplified by the fact that we assume the source sky position to be accurately known, and that we also know the arrival time of the GWs on Earth. We then describe how PNS modes are extracted from the coherent TF map to infer some of the PNS parameters using the universal relations described in~\cite{Torres:2019b}.

\subsection{Multi-detector coherent analysis}
We consider GWs propagating along $\hat{n}$, coming from a source whose location is defined by $\theta$ and $\phi$ in an Earth-fixed coordinate system. The center of the Earth is used to define the reference arrival time of the signal. In this coordinate system, the location ($\vecrk$) and orientation (given by the angles $\Psi1$ and $\Psi2$) of a detector are specified. The response of detector $k \in \{1...N\}$ to a GW with polarizations $h_{+}(t)$ and $h_{\times}(t)$ given in the wave propagation frame is 
\begin{eqnarray}\label{eq:response}
    h^{k}(t)& =& F_{+}^k(\vecrk, \theta, \phi, \psi)\, h_{+}(t-\delta t^k) \nonumber \\ 
    && + \,F_{\times}^k(\vecrk, \theta, \phi, \psi)\, h_{\times}(t-\delta t^k)
\end{eqnarray}
where $F_+^k$ and $F_{\times}^k$ are the detector response functions, and $\delta t^k$ is the difference in arrival times of the signal between the center of the Earth and the detector
\begin{equation}
    \delta t^k = \frac{\vecrk. \hat{n}}{c} ~.
\end{equation}
In each detector the data $d^{k}(t)$ are a linear combination of signal and noise,
\begin{equation}\label{eq:data}
    d^{k}(t) = n^{k}(t) + h^{k}(t) ~ ,
    %F_{+}^{k} h_{+}(t-\delta t^k) + F_{\times}^{k}  h_{\times}(t-\delta t^k)
\end{equation}
where the noise $n^k(t)$, assumed to be Gaussian and stationary, is fully characterized by the noise power spectral density $S^{k}(f)$
\begin{equation}
    < \tilde{n}^k(f) \tilde{n}^{k*}(f')> = \delta(f-f') S^{k}(f) ~ . 
\end{equation}

The PNS oscillation modes have slow and smooth evolution that show up in the TF representation of the data. For this reason, and following~\cite{Sutton:2009gi,Klimenko:2005xv}, after having time-shifted the data streams by $\delta t^k$ we consider a discrete TF representation of the data symbolized by TF pixel $\tilde{x}^k[i]$, where $i$ is a 2-dimensional index. To address Gaussian-distributed variables, all quantities in Eq.~(\ref{eq:data}) are whitened such that the discrete expression of this equation is
\begin{equation}
     \frac{\tilde{x}^{k}[i]}{\sqrt{S^k[i]}} = \frac{\tilde{n}^{k}[i]}{\sqrt{S^k[i]}} + \frac{F_{+}^{k}}{\sqrt{S^k[i]}} \tilde{h}_{+}[i] + \frac{F_{\times}^{k}}{\sqrt{S^k[i]}} \tilde{h}_{\times}[i]\,.
\end{equation}
Considering now $N$ detectors, we have a system of equations that can be written in a simple matrix form $ \mathbf{\tilde{x}} = \mathbf{\tilde{n}} + \mathbf{F} \mathbf{\tilde{h}}$,
where $\mathbf{\tilde{x}}$, $\mathbf{\tilde{n}}$ and $\mathbf{F=\{F_{+},F_{\times}\}}$ are noise-weighted quantities.
The TF pixels are coherently combined in the likelihood ratio~\cite{Flanagan:1997kp} 
\begin{equation}
    \Lambda = \frac{p(\mathbf{\tilde{x}}|\mathbf{\tilde{h}})}{p(\mathbf{\tilde{x}}|0)} ~ ,
\end{equation}
where $p(\mathbf{\tilde{x}}|\mathbf{\tilde{h}})$ is the joint probability of measuring the data $\mathbf{\tilde{x}}$ from the GW $\mathbf{\tilde{h}}$, while $p(\mathbf{\tilde{x}}|0)$ is the probability of obtaining this same data in the absence of any GW. For a set of pixels $\{i\}$,
\begin{equation}
    p(\mathbf{\tilde{x}}|\mathbf{\tilde{h}}) = \frac{1}{\sqrt{2\pi}} \exp \Bigl( -\frac{1}{2} \sum_{i} \Big|\mathbf{\tilde{x}[i]} - \mathbf{F\tilde{h}[i]}\Big|^2 \Bigl)\,,
\end{equation}
and the log-likelihood ratio is then
\begin{equation}
    {\cal{L}}=\log \Lambda =\frac{1}{2} \sum_{i} \Big\{\Big|\mathbf{\tilde{x}[i]}\Big|^2 - \Big|\mathbf{\tilde{x}[i]} - \mathbf{F\tilde{h}[i]}\Big|^2 \Big\}\,.
\end{equation}
As the sky position of the source is assumed to be known accurately, only the polarization angle $\Psi$ is a free parameter that can be arbitrarily chosen as the detector response given in Eq.~(\ref{eq:response}) is invariant under a rotation around the wave propagation axis. As demonstrated in~\cite{Klimenko:2005xv}, a particular choice of this angle allows for a better-defined antenna pattern basis, the so-called dominant polarization frame, where the detector response is maximum for the equivalent ``plus" polarization and minimal for the orthogonal polarization. In this reference frame, the antenna pattern functions $\mathbf{f_{+}}$, $\mathbf{f_{\times}}$ are orthogonal and the likelihood ratio maximized on $\mathbf{\tilde{h}}$, called the standard likelihood, has the following expression:
\begin{equation}
\label{eq:standardlikelihood}
     E_{\rm SL} = 2 {\cal L}(\mathbf{\tilde{h}_{max}}) = \sum_{i} \big\{ |\mathbf{e^{+}}\cdot\mathbf{\tilde{x}[i]}|^2 + |\mathbf{e^{\times}}\cdot\mathbf{\tilde{x}[i]}|^2 \big\} ~ ,
\end{equation}
where $\mathbf{e_{+}=f_{+}/|f_{+}|}$, $\mathbf{e_{\times}=f_{\times}/|f_{\times}|}$ and the $\cdot$ symbol denote the scalar product of one-dimensional vectors in the detector space. 

\begin{figure}[t]
 \centering
 \includegraphics[width=0.5\textwidth]{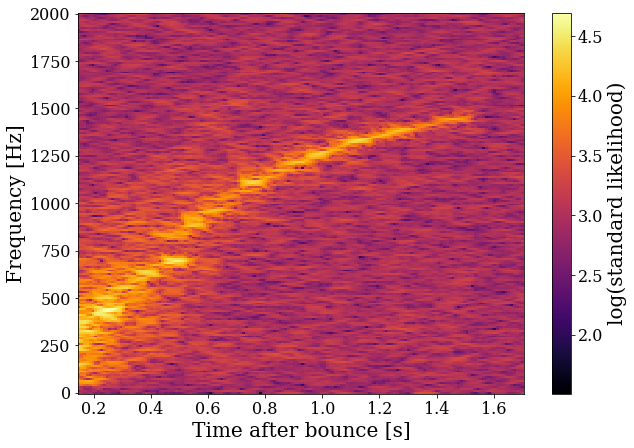}
 \caption{Time-frequency map of standard likelihood for $\sim$ \unit[2]{s} of LIGO, Virgo and KAGRA data containing the CCSN GW signal {\texttt s20} for a source at \unit[5]{kpc} in the direction RA=\unit[18.34]{h}, dec=\unit[-16.18]{\textdegree} and a GPS time of arrival on Earth $t_0$=\unit[1325048418]{s}. Time starts at the time of the core bounce.} 
 \label{fig:stdLike}
\end{figure}

In practice, for each detector we time-shift the data streams by the appropriate time delay $\delta t$, Fourier transform overlapping short-duration segments of data that have been previously Hann-windowed, calculate $\mathbf{f_{+}}$ and $\mathbf{f_{\times}}$ in the dominant polarization frame, and divide all quantities by the detector noise ASD before computing the scalar product of Eq.~(\ref{eq:standardlikelihood}). The frequency resolution of the TF maps is set by the sampling frequency of the GW data extracted from the detectors, and by the length $L$ of the segments that are Fourier transformed. The time resolution of the TF maps is inversely constrained by this same length; increasing $L$ gives better frequency resolution but at the same time degrades the time resolution. Working with overlapping segments allows to avoid this constraint. For instance, all time-frequency maps considered in this paper have been generated with time and frequency resolutions of \unit[10]{ms} $\times$ \unit[10]{Hz}.
An example of a coherent TF map obtained for the \texttt{s20} CCSN GW signal buried in noise of the LIGO-Virgo-KAGRA network is given in Figure \ref{fig:stdLike}.

\subsection{Signal tracking}
We assume that among all the possible PNS modes, the $^2g_2$-mode (we follow the nomenclature of~\cite{Torres:2019b}) carries most of the energy and that the dominant frequency forms a ridge in the spectrogram. Hereafter we refer to this mode simply as the g-mode.
In~\cite{Bizouard:2020sws} we demonstrated that a simple time-frequency method is able to track the ridge in a spectrogram taking into account the fact that the frequency monotonically increases with time. The algorithm performed well thanks to the fine frequency resolution given by the auto-regressive estimate of the local spectrum using 90\% overlapping short segments. In the coherent maps, however, the frequency of the modes is spread over more bins, making this tracking algorithm inefficient.

\begin{figure}[t]
 \centering
 \includegraphics[width=0.5\textwidth]{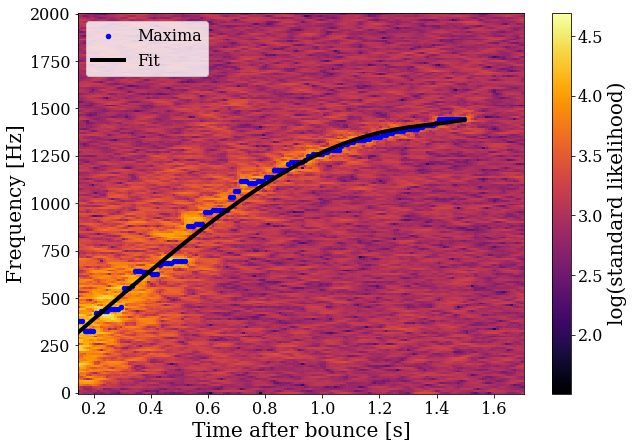}
 \caption{Application of the tracking algorithm to the standard likelihood map presented in Figure \ref{fig:stdLike} (employing the \texttt{s20} waveform buried in LIGO-Virgo-KAGRA noise). Blue circles show pixels of maximum intensity at each time index. Black dots are the results of the LASSO regression.} 
 \label{fig:tracking}
\end{figure}

Here, we propose another algorithm based on a polynomial fit of the dominant frequency. We consider for each time bin $t_i$ the pixel of maximum intensity and we record its frequency $f_i$. We fit a regularized polynomial to all those frequency values following the LASSO algorithm~\cite{Tibshirani:2015}. This method fits a given set of data $\{t_i,f_i\}$ with a polynomial of arbitrary maximal degree $p$ without the risk of over-fitting thanks to the regularization constraint which causes most monomials to be identically zero. The solution of the minimization problem are the coefficients $\{\hat{\gamma}_i\}$ that minimize
\begin{equation}
    \underset{\gamma_0,...,\gamma_p}{\min}||f_i-\sum_{i=0}^p\gamma_it^i||_2^2+\lambda\sum_{i=0}^p|\gamma_i| ~ ,
\end{equation}
where $\lambda$ is the regularization parameter and $||~||_2$ is the $\mathrm{L_2}$-norm. In the following we have used $p=10$ and $\lambda=1$. For most CCSN signals considered in this study this choice has resulted in less than five non-zero $\gamma_i$ coefficients.
To handle outlier data points, the regularized polynomial regression is performed a second time after removing the points that are more distant from the first fit than the root-mean-square deviation. By interpolation, we obtain a function $f(t)$ which represents the time evolution of the g-mode frequency as displayed in Figure \ref{fig:tracking}.

\begin{figure*}[ht]
 \centering
 \begin{tabular}{cc}
 \includegraphics[width=0.5\textwidth]{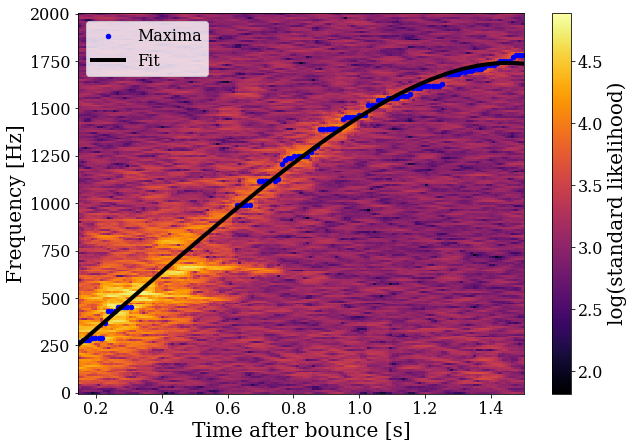} &
 \includegraphics[width=0.5\textwidth]{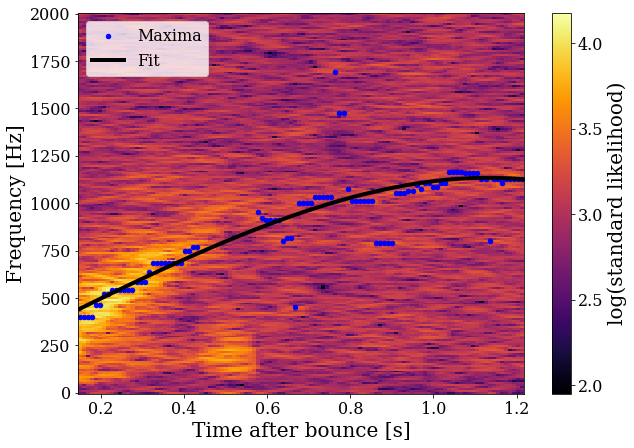}
 \end{tabular}
 \caption{{\bf Left:} Time-frequency map of standard likelihood for $\sim$ \unit[2]{s} of LIGO, Virgo and KAGRA data containing CCSN GW signal {\texttt s25} for a source at \unit[2]{kpc} in the direction RA=\unit[18.34]{h}, dec=\unit[-16.18]{\textdegree} and a GPS time of arrival on Earth $t_0$=\unit[1325048418]{s}. Time starts at the time of the core bounce. The pixels of maximum energy with frequencies between \unit[500]{Hz} and \unit[700]{Hz} do not appear on the map as they are not taken into account by the tracking algorithm. {\bf Right:} As the left panel but for waveform \texttt{s15--3Dp}. Here it is the pixels between \unit[0.45]{ms} and \unit[0.6]{ms} that are not taken into account for tracking.}
 \label{fig:s25and3D}
\end{figure*}

As this tracking technique considers the pixels of maximum intensity over the entire time-frequency map, it may be affected by some features of the map that deviate from the assumption of a single, continuous track. To overcome this, an option is added to define a certain area of the map that will not be taken into account in the LASSO regression. For example, in the case of the waveforms with the most massive progenitors \texttt{s25} and \texttt{s40}, the linear analysis of~\cite{Torres:2018,Torres:2019a} has shown that the $^2g_2$ mode interacts with the $^2g_3$ mode in an avoided crossing a few \unit[100]{ms} after bounce. The signature of this additional mode is either a gap in the dominant g-mode emission and/or a down-going secondary feature in the spectrograms (see left panel of Figure \ref{fig:s25and3D}). It is then possible to define a frequency band around this avoided crossing ($f \in [500,700]$Hz) that is left out of the fit.
In the case of the two waveforms extracted from the 3D simulation, we observe a very energetic low-frequency component (see right panel of Figure \ref{fig:s25and3D}) which is a signature of the standing accretion shock instability (SASI)~\citep[see, e.g.,][]{Murphy:09, Cerda:2013, mueller:13gw, Yakunin:2015wra, Kuroda:2016, Andresen:2017, powell:2019, Radice:2018usf}. For this model we define the time band in which the SASI is most energetic ($t \in [0.45, 0.6]$s) and that is ignored for the fit. 

In the context of this study there is no generic method for selecting a region of the TF map that should be ignored for the tracking. This would require an algorithm that could track simultaneously different modes and it has not been implemented in our pipeline. In the event of an actual CCSN GW detection, we would have to eliminate `by eye' the features that are clearly not related to the g-mode (such as the low-frequency SASI emission).

\subsection{Parameter estimation}
The strategy to estimate the time evolution of the ratio $M_{\rm PNS}/R_{\rm PNS}^2$ is similar to what we proposed in~\cite{Bizouard:2020sws}. We first build a model that describes the ratio as function of the frequency $f$ based on the subset of 18 out of all the 25 1D simulations of \cite{Torres:2019b} that were generated with the \texttt{AENUS-ALCAR} code. We parametrized the ratio and frequency with a cubic polynomial regression with heteroscedastic errors
\begin{equation}
\begin{split}
    r&=\beta_1f+\beta_2f^2+\beta_3f^3+\epsilon\\
    \log\sigma&=\alpha_0+\alpha_1f+\alpha_2f^2 ~ ,
\end{split}
\label{eq:model}
\end{equation}
where $\epsilon$ is a zero-mean Gaussian error term with a frequency-dependent variance $\sigma$.
Because we are considering the same set of 1D simulations, the fit of the model obtained using the R-package \texttt{lmvar}~\cite{lmvar:2019} is providing the same $\alpha$ and $\beta$ coefficients given in Table 2 of~\cite{Bizouard:2020sws}.
We can then inject the interpolated frequency of the high-frequency ridge inside Eq.~(\ref{eq:model}) to obtain an estimate of the time evolution of the ratio $r(t)$ as well as 95\% confidence intervals.

\section{Observational prospects}
\label{sec:results}
\begin{figure}
 \centering
 \includegraphics[width=0.5\textwidth]{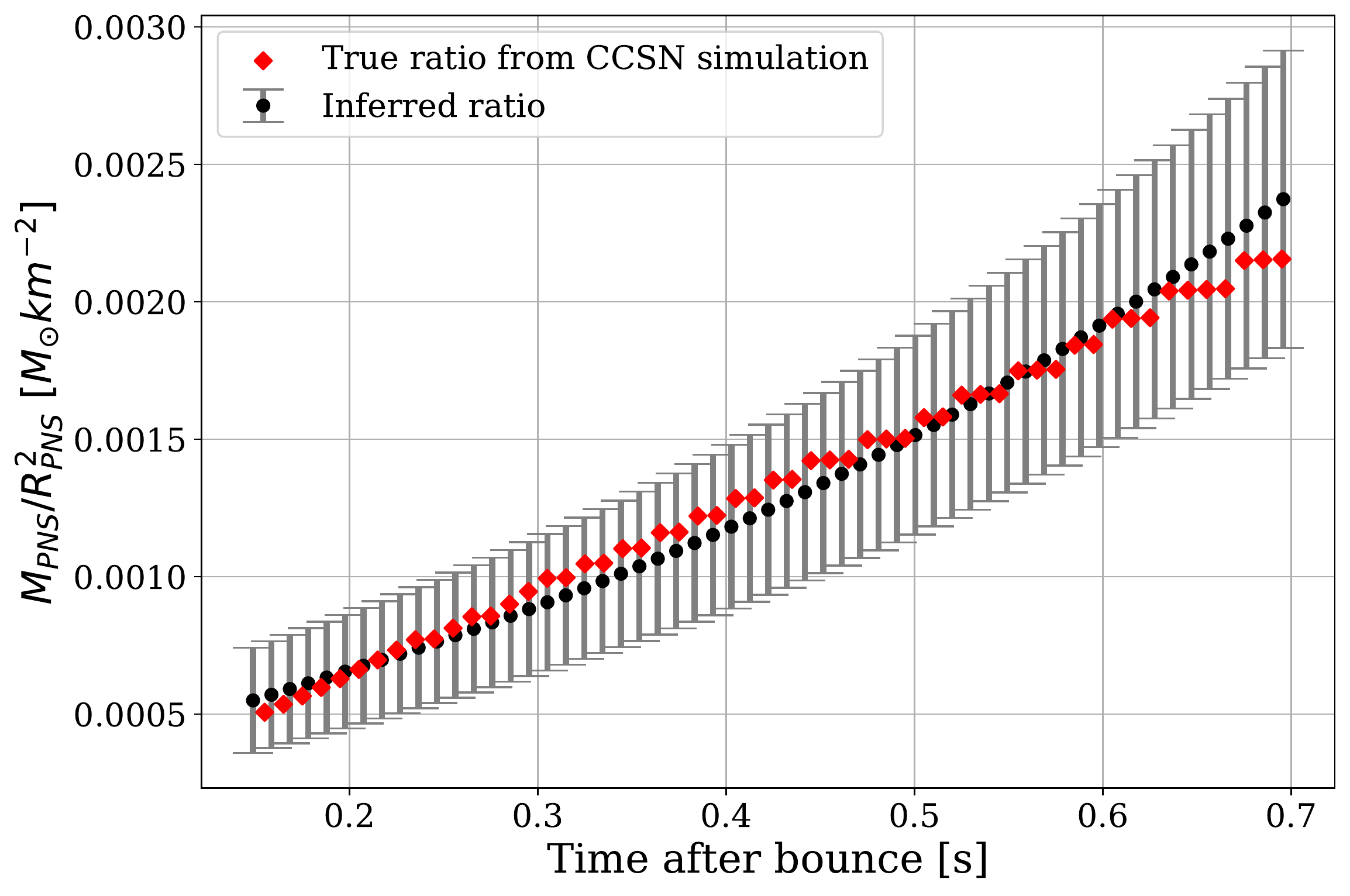}
 \caption{Application of our parameter inference algorithm on the same example used in Figure \ref{fig:stdLike}. The source is simulated at \unit[5]{kpc} in the direction RA=\unit[18.34]{h}, dec=\unit[-16.18]{\textdegree} and detected with the HLVKA network. Red markers show the time evolution of the ratio $M_{\rm PNS}/R_{\rm PNS}^2$ as given by the simulation from which the 2D waveform \texttt{s20} has been extracted. Black markers show the time evolution of the ratio estimated from the high-frequency GW signal. The grey band shows the 95\% confidence intervals. }
 \label{fig:ratio}
\end{figure}

To assess the performance of the inference method, we use the CCSN waveforms presented in Section~\ref{sec:simulations} in different detector network configurations and source locations. In all cases, we reconstruct the time evolution of the ratio $r(t)$ and compare it with the evolution of the ratio $M_{\rm PNS}/R_{\rm PNS}^2$ provided by the CCSN simulation from which the waveform was extracted. We define the coverage as the fraction of the true values that lie within the 95\% confidence interval of the inferred values. This index is used to quantify the precision of the ratio reconstruction. Figure \ref{fig:ratio} shows the time evolution of $M_{\rm PNS}/R_{\rm PNS}^2$ in the CCSN simulation \texttt{s20} along with the inferred ratio obtained with the GW waveform injected to the data of the LIGO-Virgo-KAGRA network for a source located at \unit[5]{kpc} in the direction of the Sagittarius constellation (RA=\unit[18.34]{h}, dec=\unit[-16.18]{\textdegree}). In this case all the true ratios (in red) are inside the 95\% band of the reconstructed ratios (in black) and the coverage is equal to 1.

In the following we systematically report the median value of the coverage for 100 different noise realisations in the detectors.

\subsection{Inference distance}
\label{sub:obs}
We first compare the evolution of the coverage as function of the distance to the source when using only the two LIGO interferometers (HL) with the case of the full LIGO-Virgo-KAGRA network (HLVKA). We consider a fixed sky position for the source inside the Sagittarius constellation, as Galactic CCSNe are expected from the Galactic plane. We set this sky position to RA=\unit[18.34]{h}, dec=\unit[-16.18]{\textdegree} and we vary the distance to the source. As the antenna patterns of one interferometer depend on the direction of propagation of the GWs with respect to the orientation of its arms, the Earth rotation modulates the signal amplitude and thus affects the reconstruction quality. We consider two cases, one favourable for the HLVKA network and the other one unfavourable based on the value of the equivalent antenna pattern of a network composed of $N$ detectors defined as
\begin{equation}
    F_{\rm eq}=\sqrt{\frac{1}{N}\sum_{k\in {\rm network}}{({F_{+}^{k}(\overrightarrow{\Omega}, t_0)}^2 + {F_{\times}^k(\overrightarrow{\Omega}, t_0)}^2)}} ~ ,
\end{equation}
where $\overrightarrow{\Omega}$ is the sky position of the source and $t_0$ is the arrival time of the GWs at the center of the Earth.
The favourable case corresponds to an arrival time that maximizes $F_{\rm eq}$ for the fixed sky position, namely $F_{\rm eq} \sim 0.53$ for $t_0$=\unit[1325052478]{s}. Similarly, the unfavourable case corresponds to the smallest value of $F_{\rm eq} \sim 0.38$ obtained for $t_0$=\unit[1325077869]{s}. We point out that this definition of favourable and unfavourable cases does not take into account the different sensitivities of the interferometers but only their antenna patterns. Therefore, the favourable case described here is not necessarily the best possible scenario but only serves as a comparison with the second case which is a less advantageous scenario.

\begin{figure*}[ht]
 \begin{tabular}{cc}
 \includegraphics[width=0.5\textwidth]{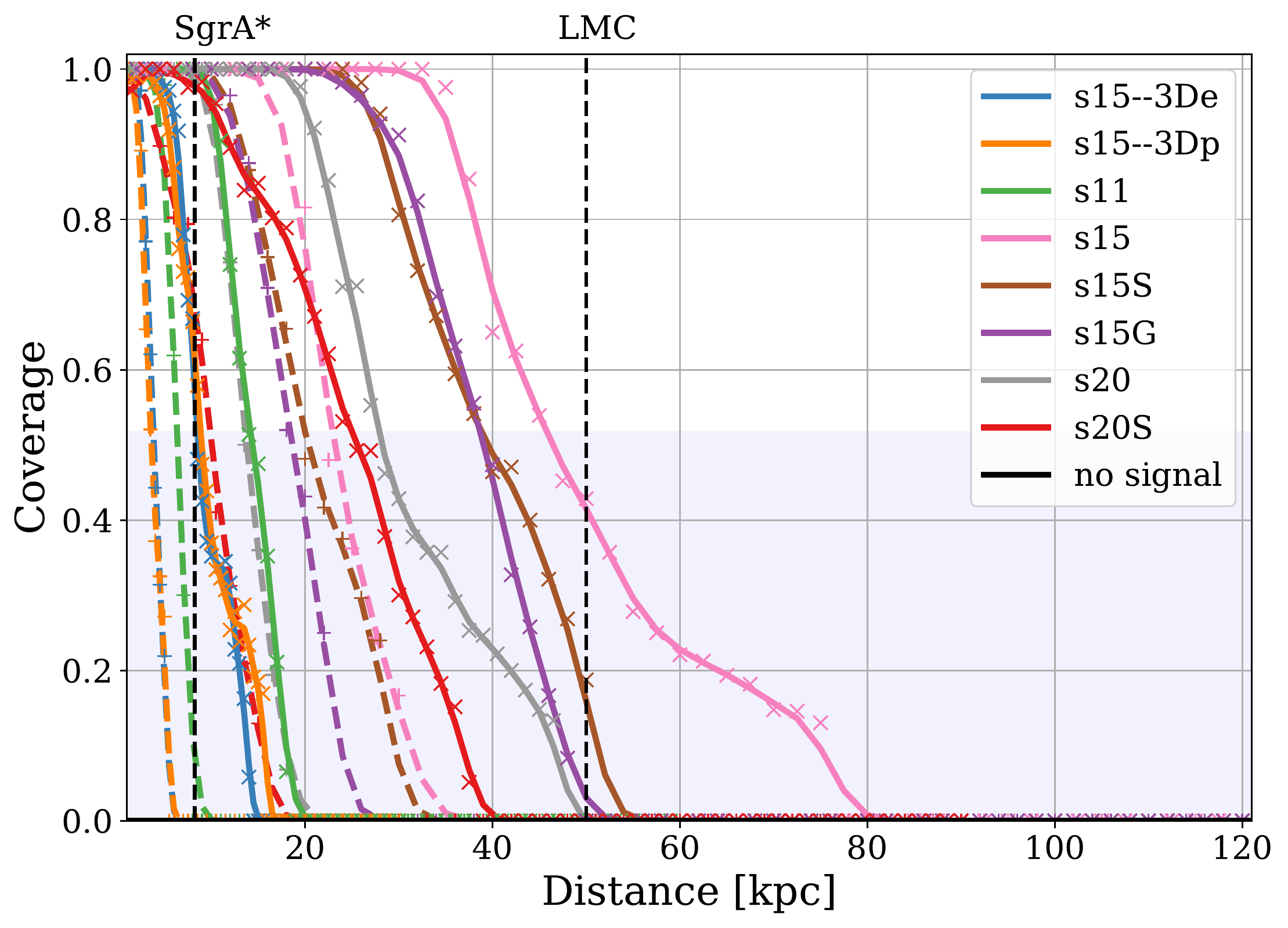}&
 \includegraphics[width=0.5\textwidth]{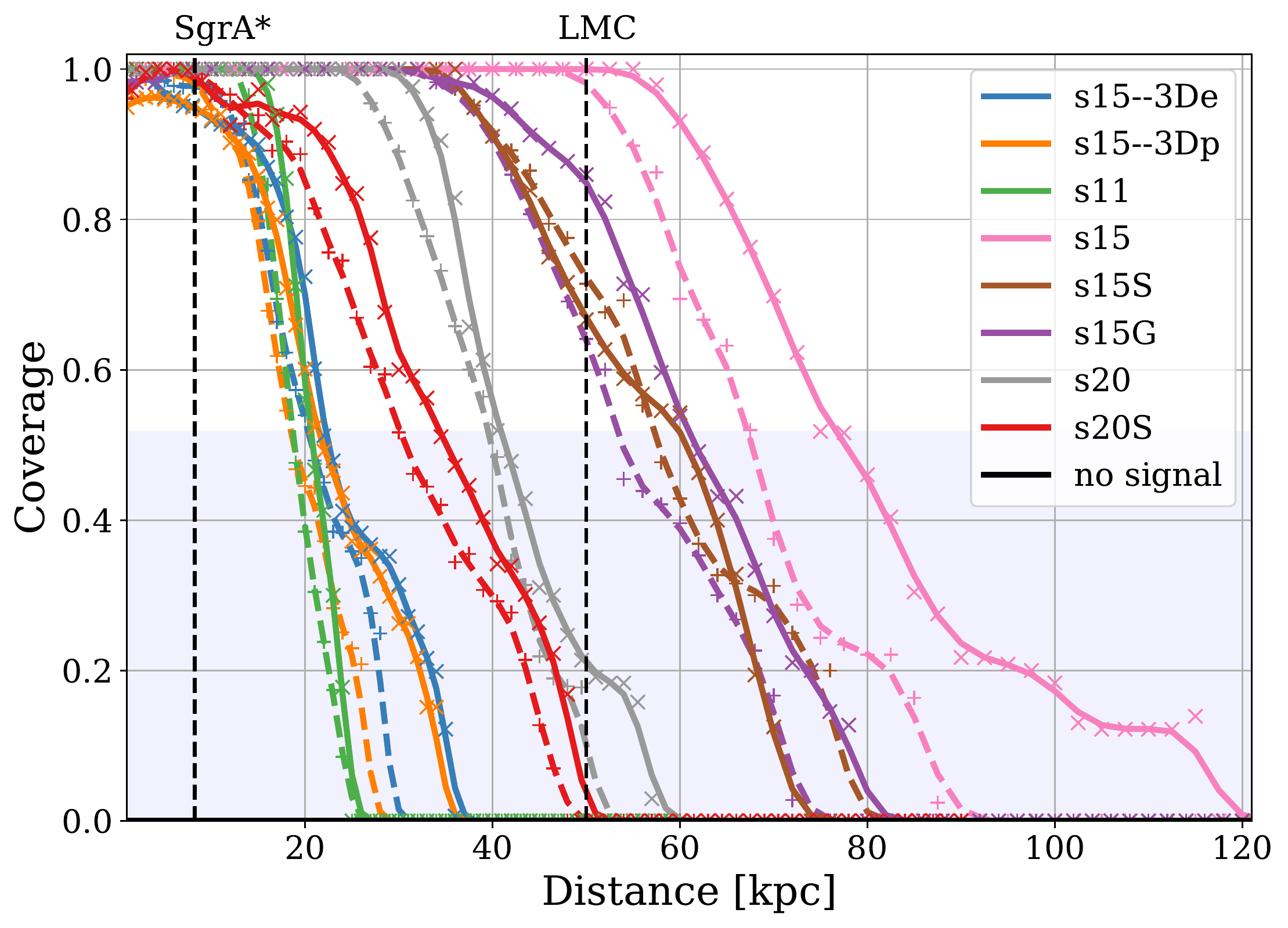}
 \\
 \includegraphics[width=0.5\textwidth]{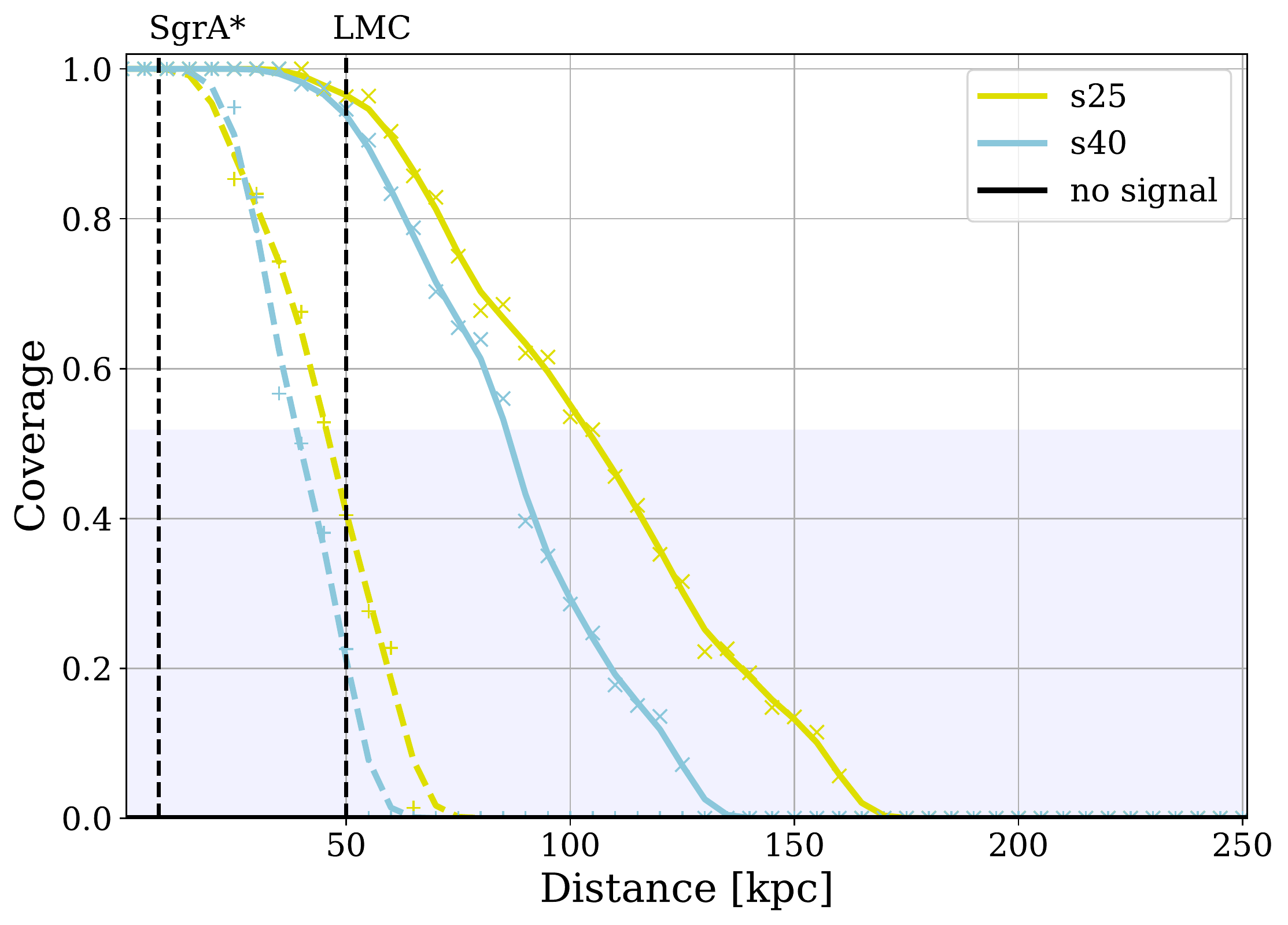}&
 \includegraphics[width=0.5\textwidth]{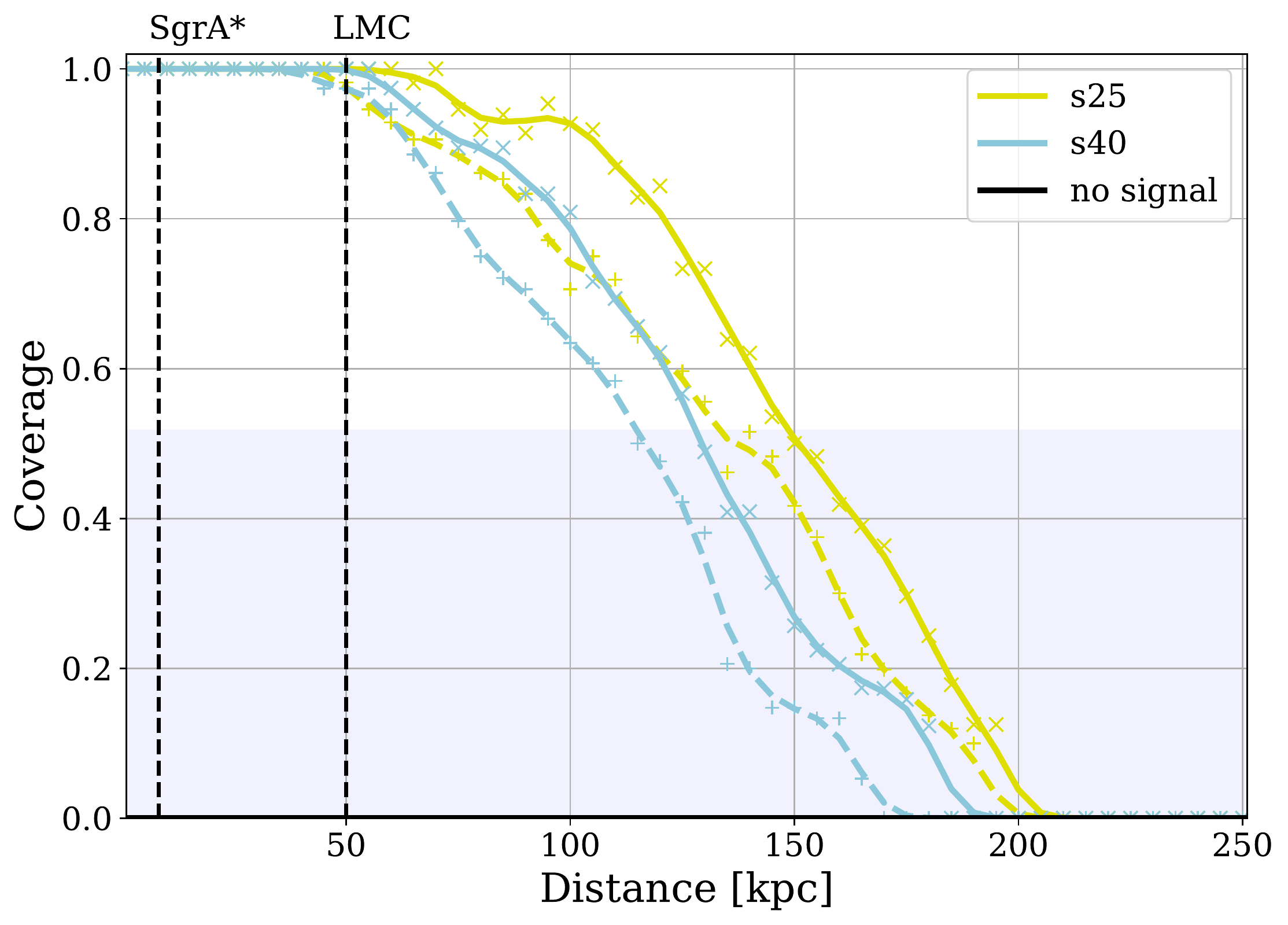}
 \end{tabular}
 \caption{Evolution of the median coverage as a function of distance to the source for eight 2D and two 3D CCSN waveforms reconstructed with two second-generation GW detector networks. The sky position of the source is set in the direction of the Sagittarius constellation, namely RA=\unit[18.34]{h} and dec=\unit[-16.18]{\textdegree}. For each waveform a solid line shows the smoothing splines of the median coverage when the signal is injected in the HLVKA network while a dashed line is for the HL network. The left (right) panel corresponds to a GW signal arrival time which is unfavourable (favourable) in the HLVKA network. In the unfavourable (favourable) case the equivalent antenna pattern is $F_{\rm eq}\sim 0.38$ ($F_{\rm eq}\sim 0.53$). The upper panel displays the results obtained with the two waveforms extracted from the 3D simulation and the first six waveforms from our test set of 2D simulations, while the lower panel shows the remaining two 2D waveforms \texttt{s25} and \texttt{s40}. The ``no signal'' line shows the median coverage in the case where no signal is present in the data (source at a distance of \unit[$10^6$]{kpc}). It is always strictly equal to 0 for both the unfavourable and favourable cases. The blue band represents the $\mathrm{95^{\rm th}}$ percentile of coverage over 1000 different noise realisations for the ''no signal'' case.}
 \label{fig:perf_2G}
\end{figure*}

Figure~\ref{fig:perf_2G} shows the evolution of the coverage as a function of the distance to the source for the two different arrival times considered. In the absence of signal, the median coverage is null. The blue band on Figure~\ref{fig:perf_2G} represents the $95^{\rm th}$ percentile of the coverage obtained for 1000 simulations with a waveform injected at a distance of \unit[$10^6$]{kpc} (`no signal' case). The upper limit of the band, $\sim 0.53$, is an indication that for coverage values below this value the ratio reconstruction should not be trusted. 
In the rest of the paper we will consider that the ratio is well reconstructed for coverage values above $0.8$. The two 3D waveforms and six of the eight 2D waveforms exhibit the same monotonic behaviour with a coverage equal to unity at close distances followed by a drop and a convergence to zero at large distances. The quality of the ratio reconstruction with the HLVKA network (solid lines) is always better than with the HL network only (dotted lines). This is the case for all waveforms with the exception of \texttt{s15S} for which there is no clear difference in the favorable case. This confirms that having a large multi-detector network increases the capabilities to accurately estimate the PNS physical parameters. In the favourable case (right panels), the ratio is reconstructed with a good precision (with a coverage greater to 0.8) for distances up to Sagittarius A\*, the centre of the Milky Way, including the low mass progenitor simulations. For very energetic events, it would even be possible to infer the physical parameters of the PNS at distances up to the Large Magellanic Cloud in some cases (e.g.~\texttt{s15}, \texttt{s25} and \texttt{s40}, the last two shown in the bottom panel of Figure~\ref{fig:perf_2G}).

\begin{figure}[ht]
 \includegraphics[width=0.5\textwidth]{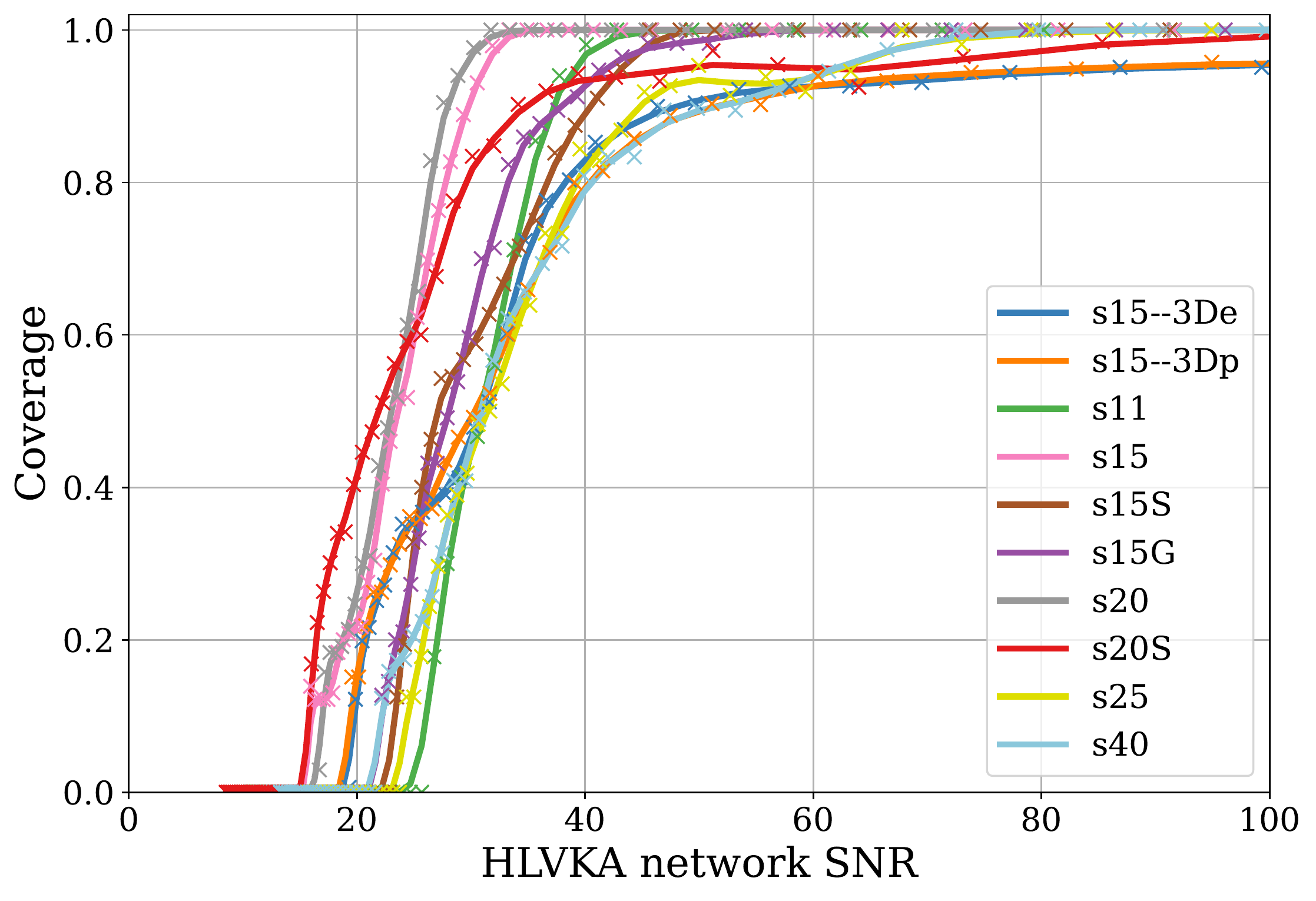}
 \caption{Evolution of the median coverage as a function of network signal-to-noise ratio (SNR) for the 10 CCSN waveforms injected in the network of second generation GW detectors HLVKA. The sky position of the source is set in the direction RA=\unit[18.34]{h}, dec=\unit[-16.18]{\textdegree}, inside the Sagittarius constellation.} 
 \label{fig:perf_2G_SNR}
\end{figure}

\begin{table}
    \centering
    \begin{tabular}{c|c|c|c|c|c}
    \hline
        \multirow{3}{*}{Waveform} & \multirow{3}{*}{Energy [$M_{\odot}c^2$]} & \multicolumn{4}{c}{Distance reach [kpc]} \\
        \cline{3-6}
         & & \multicolumn{2}{c|}{\textit{ unfavourable}} & \multicolumn{2}{c}{\textit{favourable}} \\
         \cline{3-6}
         & & HL & HLVKA & HL & HLVKA \\
        \hline
        \texttt{s15--3De} & 1.75e-9 & 2.9 & 6.9 & 15.5 & 18.1 \\
        \texttt{s15--3Dp} & 7.66e-10 & 2.7 & 6.3 & 15.0 & 17.0 \\
        \texttt{s11} & 3.20e-9 & 5.5 & 11.6 & 16.3 & 18.4 \\
        \texttt{s15} & 3.08e-8 & 20.1 & 38.2 & 58.4 & 66.1 \\
        \texttt{s15S} & 8.38e-9 & 15.1 & 30.2 & 45.8 & 44.9 \\
        \texttt{s15G} & 6.85e-9 & 14.9 & 32.4 & 44.3 & 52.4 \\
        \texttt{s20} & 1.03e-8 & 11.6 & 23.1 & 32.3 & 36.3 \\
        \texttt{s20S} & 2.68e-9 & 6.4 & 16.7 & 21.4 & 26.4 \\
        \texttt{s25} & 9.26e-8 & 31.8 & 71.8 & 92.7 & 122.0 \\
        \texttt{s40} & 4.51e-8 & 30.5 & 63.7 & 74.8 & 100.5 \\
        \hline
    \end{tabular}
    \caption{Gravitational-wave energy of all the waveforms considered in this study and distance up to which the ratio is reconstructed with a good accuracy (coverage greater than $0.8$) for two different arrival times of the GW and in two different network configurations (HL and HLVKA).}
    \label{tab:perf}
\end{table}

The distances up to which the coverage remains greater than 0.8 are reported in Table~\ref{tab:perf} for all waveforms and both arrival times.
As expected, the distance reach approximately scales with the total energy carried by the GWs. The size of the detector network also matters, especially when none of the detectors' orientation is optimal for the source. In the favourable case there is an average $\sim$ 17\% improvement of the distance reach with the HLVKA network compared to HL only, while an improvement of $\sim$ 118\% in the unfavourable case is observed. Remarkably, for almost all progenitor masses considered, it will be possible to infer the physical parameters of the PNS for a source within the Milky Way.

Another way to look at the performance of the inference method is to represent the coverage as function of the signal-to-noise ratio (SNR) in the network of detectors
\begin{equation}
    {\rm SNR}^{\rm net}(w,d,\overrightarrow{\Omega},t_0) = \sqrt{\sum_{k}{{{\rm SNR}_k(w,d,\overrightarrow{\Omega}, t_0)}^2}}
\end{equation}
where ${\rm SNR}_k(w,d,\overrightarrow{\Omega}, t_0)$ is the matched filter SNR of waveform $w$ in detector $k$ for a source at a distance $d$, with a sky position $\overrightarrow{\Omega}$ and a time of arrival of the GWs at the center of Earth $t_0$. Using the signal injections of the favourable case described above, we show in Figure~\ref{fig:perf_2G_SNR} the evolution of the coverage as a function of the HLVKA network SNR. We observe that the performance of our method depends only slightly on the waveform considered and a network SNR between $27$ and $40$ is required to make a reliable inference on the PNS physical parameters. The lower limit of the SNR is obtained for waveforms in which the majority of the GW energy is carried out by the $^2g_2$ mode (for instance waveform \texttt{s20}). This ${\rm SNR}^{\rm net}$ range corresponds to an individual LIGO Hanford SNR range of $17 - 22$ which is comparable to the detection performance (at 50\% efficiency) of the coherent Waveburst CCSN search pipeline reported in~\cite{Szczepanczyk:2021bka}.

\subsection{Effect of the number of detectors in the network}
\label{sub:net}

\begin{figure*}[ht]
 \begin{tabular}{cc}
 \includegraphics[width=0.5\textwidth]{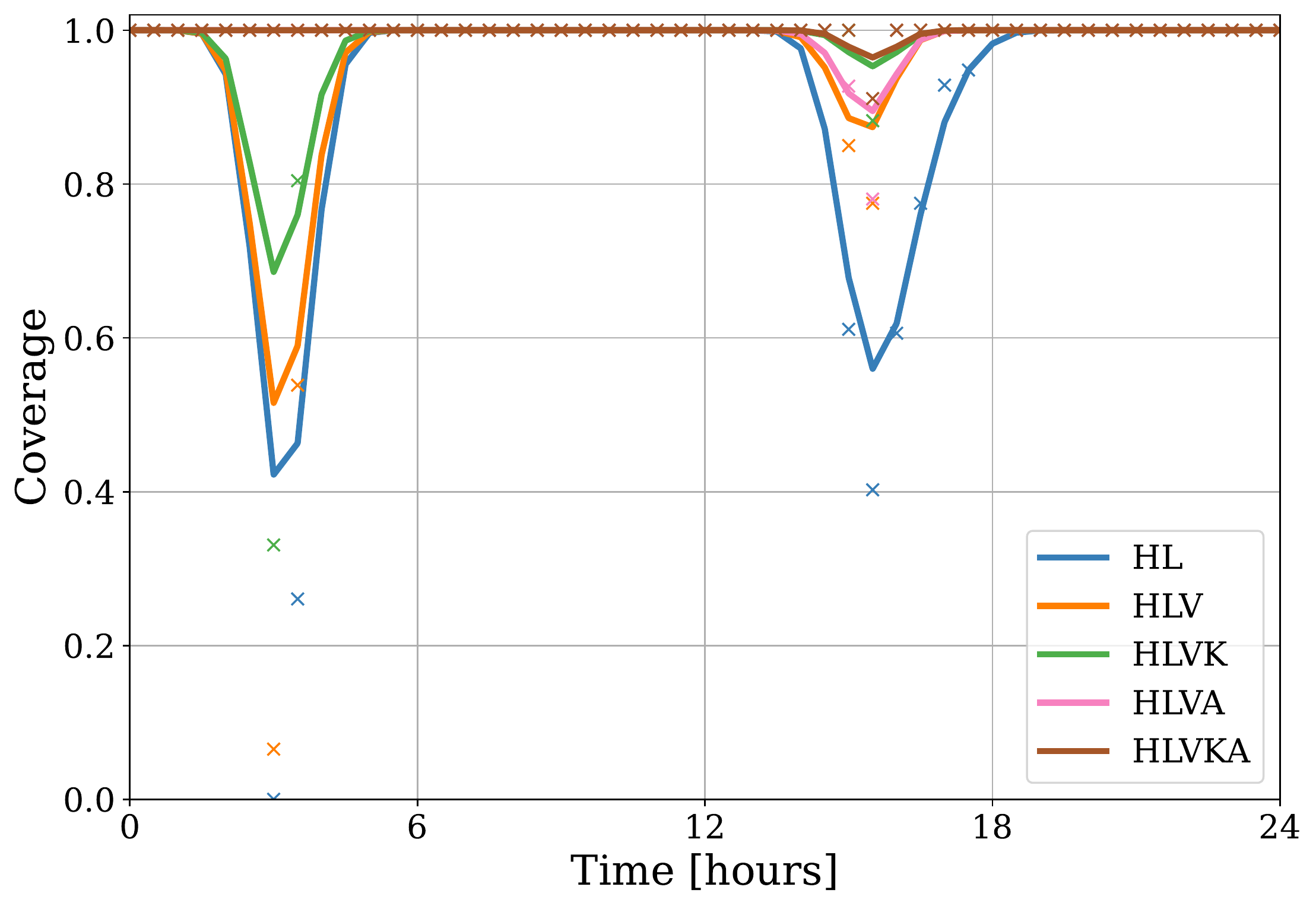}&
 \includegraphics[width=0.5\textwidth]{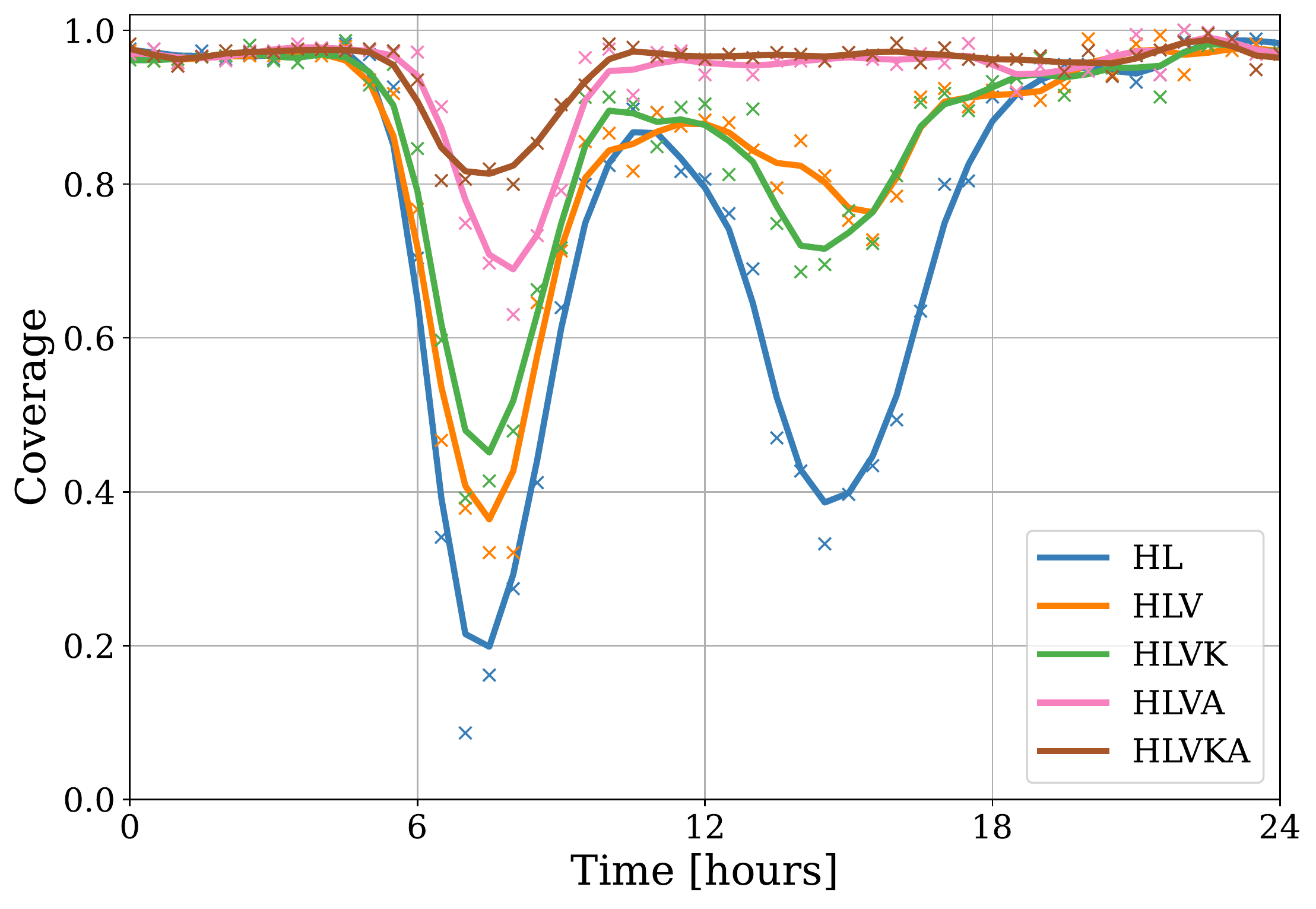}
 \end{tabular}
 \caption{Evolution over 24 hours of the median coverage for a CCSN source located at the center of the Milky Way (RA=$\mathrm{17^h45^m}$, dec=$\mathrm{-29^{\circ}00^{'}}$, d=\unit[8.2]{kpc}) and for different arrival times of the GWs. The data employed correspond to the \texttt{s20} (left) and \texttt{s15--3De} (right) waveforms. Each color represents a different network configuration, HL, HLV, HLVK, HLVA and HLVKA, as indicated in the legends.} 
 \label{fig:galacenter}
\end{figure*}

We now take a closer look at the effect of the Earth rotation on the precision of the ratio reconstruction and at the role played by the addition of each interferometer in the network. We fix a sky position for the source at the Galactic Center (RA=$\mathrm{17^h45^m}$, dec=$\mathrm{-29^{\circ}00^{'}}$, d=\unit[8.2]{kpc})~\cite{GalacticCenter:2021}. To simulate the rotation of the Earth we vary the arrival time of the GWs on Earth over a 24-hour period. For all the waveforms considered the GW signal is injected in one of the following network configurations HL, HLV, HLVK, HLVA or HLVKA. In the left-hand panel of Figure \ref{fig:galacenter}, the median value of the coverage obtained with waveform \texttt{s20} is shown. Considering the HL network formed by the almost co-aligned LIGO Hanford and LIGO Livingston detectors as the reference network, the addition of Virgo improves the reconstruction when HL is already performing at its best ($t=$\unit[6]{h} and $t=$\unit[12]{h}) and covers a time window where HL is totally blind to a signal (between $t=$\unit[16]{} and $t=$\unit[19]{h}). The LIGO Aundha detector allows for the optimal reconstruction of the signal between $t=$\unit[6]{h} and $t=$\unit[9]{h}, during which none of the other detectors are observing well. On the other hand, it seems that the KAGRA detector is barely contributing to the overall performance. This is partly due to the specific waveform model we use in this example as when using a waveform such as \texttt{s20} which has been obtained from a 2D CCSN simulation, only the plus polarization is available and in that case the cross antenna pattern of the interferometers do not play any role in the analysis. This happens to disfavour KAGRA as its cross antenna pattern is the strongest when the two LIGO and Virgo are in a blind spot. To check this we plot on the right-hand panel of Figure \ref{fig:galacenter} the coverage obtained with the 3D waveform \texttt{s15--3De}, where the two polarizations are active, over a 24-hour period. The results obtained, however, do not show any clear evidence of the effect of the cross polarization. This comparison seems to indicate that the lower sensitivity of KAGRA is ultimately the determining factor for its minor contribution to a network of second-generation detectors.

Table \ref{tab:galacenter} reports the fraction, averaged over 24 hours, of the coverage greater than 0.8 for each waveform and each network configuration considered. Adding Virgo to the HL network consistently gives better results with an average improvement of $\sim 25\%$ for the 3 less energetic waveforms \texttt{s15--3De}, \texttt{s15--3Dp} and \texttt{s11}. This average improvement increases to $\sim 51\%$ when using HLVKA network compared to HL only.

\begin{table}[ht]
    \centering
    \begin{tabular}{c|c|c|c|c|c}
    \hline
        Waveform & HL & HLV & HLVK & HLVA & HLVKA \\
        \hline
        \texttt{s15-3De} & 0.63 & 0.77 & 0.77 & 0.90 & 0.98 \\
        \texttt{s15-3Dp} & 0.67 & 0.85 & 0.77 & 0.92 & 0.94 \\
        \texttt{s11} & 0.52 & 0.65 & 0.65 & 0.81 & 0.81 \\
        \texttt{s15} & 1.0 & 1.0 & 1.0 & 1.0 & 1.0 \\
        \texttt{s15S} & 0.96 & 0.98 & 0.98 & 1.0 & 1.0 \\
        \texttt{s15G} & 0.96 & 0.98 & 1.0 & 1.0 & 1.0 \\
        \texttt{s20} & 0.88 & 0.94 & 0.98 & 0.98 & 1.0 \\
        \texttt{s20S} & 0.60 & 0.73 & 0.75 & 0.90 & 0.90 \\
        \texttt{s25} & 1.0 & 1.0 & 1.0 & 1.0 & 1.0 \\
        \texttt{s40} & 1.0 & 1.0 & 1.0 & 1.0 & 1.0 \\
        \hline
    \end{tabular}
    \caption{Fraction of the coverage greater than 0.8 for arrival times of the GWs spanning a 24-hour period with different network configurations. For each arrival time and each network configuration the CCSN is simulated at the center of the Milky Way (RA=$\mathrm{17^h45^m}$, dec=$\mathrm{-29^{\circ}00^{'}}$, d=\unit[8.2]{kpc}).}
    \label{tab:galacenter}
\end{table}

\subsection{Results with third generation detectors}
\label{sub:3G}
As it has been discussed in Section \ref{sub:net}, second-generation ground-based detectors will be able to accurately reconstruct the ratio $M_{\rm PNS}/R_{\rm PNS}^2$ for a CCSN located at a distance of up to a few tens of kpc. This limit depends mainly on the sensitivities of the interferometers and constrains the observational prospects to the Milky Way only. With the construction of the third-generation ground-based detectors Einstein Telescope~\cite{Punturo:2010zza} and Cosmic Explorer~\cite{reitze2019cosmic}, it will be possible to explore a larger volume of the Universe (see Figure \ref{fig:PSDs}). In the case of ET, the three interferometers\footnote{The triangular configuration of the Einstein Telescope has actually 2 interferometers in each of triangle sides but for simplicity we assume they are equivalent to a single one.} are combined in the likelihood function such that ET can be used alone to apply the coherent analysis method.

\begin{figure*}[ht]
    \begin{tabular}{cc}
    \includegraphics[width=0.5\textwidth]{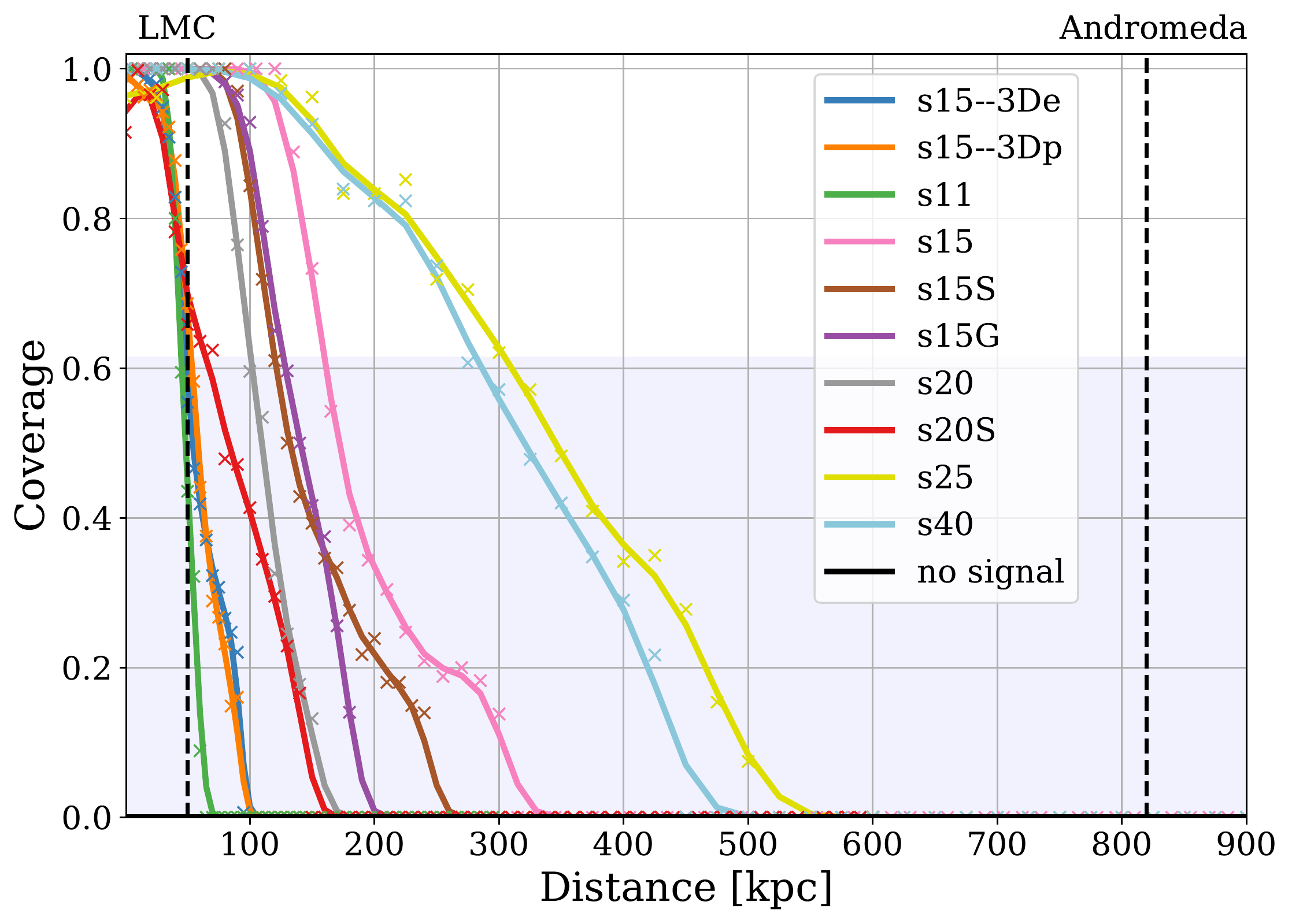}&
    \includegraphics[width=0.5\textwidth]{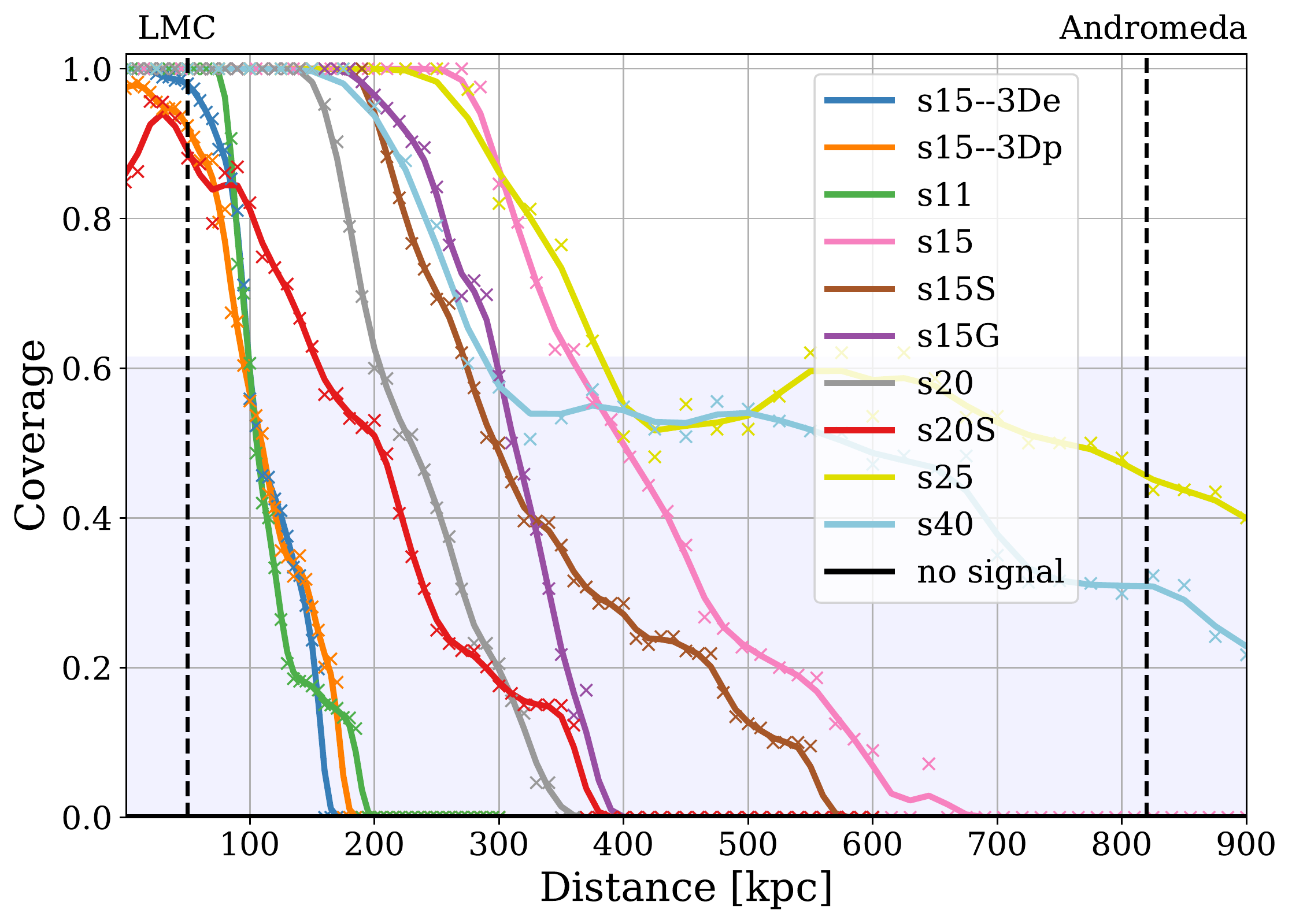}
    \end{tabular}
    \caption{Evolution of the median coverage as a function of the distance to the source for eight 2D and two 3D CCSN waveforms reconstructed with a network of third-generation GW detectors composed of ET and two CE interferometers (ET-CE20-CE40). The sky position of the source is set in the direction of the Andromeda Galaxy (RA=\unit[0.71]{h}, dec=\unit[41.27]{\textdegree}). The left (right) panel corresponds to an unfavourable (favourable) GW signal arrival time. In the unfavourable (favourable) case the equivalent '+' antenna pattern is $F^{+}_{\rm eq}\sim 0.208$ ($F^{+}_{\rm eq}\sim 0.598$).
    The ``no signal'' line shows the median coverage in the case where no signal is present in the data (source at a distance of \unit[$10^6$]{kpc}). It is strictly zero for both the unfavourable and favourable cases. The blue band represents the $\mathrm{95^{\rm th}}$ percentile of coverage over 1000 different noise realisations for the ``no signal'' case.} 
    \label{fig:perf_3G}
\end{figure*}

To assess the gain brought to the PNS inference by third-generation detectors, we repeat the analysis described in Section \ref{sub:obs} with a fixed source position set in the direction of the Andromeda Galaxy (RA=\unit[0.71]{h}, dec=\unit[41.27]{\textdegree})~\cite{Andromeda:2021}. For that sky location and for a network composed of three detectors (ET-CE20-CE40) we define again two cases corresponding to two different times of arrival of the GWs (favourable case $t_0$=\unit[1325113218]{s}, $F^{+}_{\rm eq}\sim 0.60$ and unfavourable case $t_0$=\unit[1325062818]{s}, $F^{+}_{\rm eq}\sim 0.21$). The results obtained are shown in Figure \ref{fig:perf_3G}. The time evolutions of the coverage with the distance are similar to those obtained with the second-generation detectors. However, the typical distance range is now one order of magnitude larger. In the favourable case the coverage remains greater than 0.8 up to the distance of the Large Magellanic Cloud for all the eight waveforms used. Therefore, with this new generation of detectors we will be able to observe the GWs and infer physical parameters for a CCSN that takes place in one of the dwarf galaxies around the Milky Way. According to our findings, unless the GW signal emitted by a CCSN is much more energetic than expected, the Andromeda galaxy remains out of reach even for third-generation detectors. This observational limitation is consistent with the results of \cite{Srivastava_2019} in which the authors show that even with optimized third generation interferometers, the detection range remains bound to the Milky Way and its satellite dwarf galaxies.

\begin{table*}[ht]
    \centering
    \begin{tabular}{c|c|c|c||c|c|c}
    \hline
        Galaxy name & Distance [kpc] & RA [h] & dec [\textdegree] & CE20-CE40 & ET &  CE20-CE40-ET \\
        \hline
        Sagittarius & 20 & 18.91753 & -30.47833 & 0.96 & 0.92 & 1.0 \\
        Segue 1 & 23 & 10.11756 & 16.07361 & 0.92 & 0.92 & 1.0 \\
        Tucana III & 25 & 23.94333 & -59.6 & 0.96 & 0.83 & 1.0 \\
        Hydrus I & 27.6 & 2.49261 & -79.3089 & 0.92 & 0.88 & 1.0 \\
        Carina III & 27.8 & 7.642 & -57.89972 & 0.88 & 0.92 & 1.0 \\
        Triangulum II & 30 & 2.22150 & 36.17844 & 0.88 & 0.88 & 1.0 \\
        Reticulum II & 32 & 3.59503 & -54.04917 & 0.88 & 0.84 & 1.0 \\
        Ursa Major II & 34.7 & 8.85833 & 63.13 & 0.88 & 0.84 & 1.0 \\
        Segue 2 & 35 & 2.32111 & 20.17528 & 0.67 & 0.84 & 1.0 \\
        Carina II & 36.2 & 7.60711 & -57.99917 & 0.88 & 0.79 & 1.0 \\
        Coma Berenices & 42 & 12.44972 & 23.90417 & 0.79 & 0.79 & 1.0 \\
        Bo\"otes II & 42 & 13.96667 & 12.85000 & 0.79 & 0.83 & 1.0 \\
        Willman 1 & 45 & 10.82250 & 51.05 & 0.79 & 0.79 & 1.0 \\
        Bo\"otes III & 50 & 13.95206 & 26.775 & 0.79 & 0.79 & 1.0 \\
        Large Magellanic Cloud & 50 & 5.39294 & -69.75611 & 0.75 & 0.75 & 1.0 \\
        Tucana II & 58 & 22.86531 & -58.56889 & 0.75 & 0.67 & 0.92 \\
        Small Magellanic Cloud & 60 & 0.87722 & -72.80028 & 0.67 & 0.71 & 1.0 \\
        Ursa Minor & 60 & 15.15316 & 67.21436 & 0.71 & 0.71 & 0.92 \\
        Bo\"otes & 66 & 14.0000 & 14.5000 & 0.71 & 0.71 & 0.92 \\
        Draco & 80 & 17.33732 & 57.92122 & 0.63 & 0.63 & 0.83 \\
        Sculptor & 84.3 & 1.00261 & -33.70889 & 0.67 & 0.58 & 1.0 \\
        Horologium I & 87 & 2.92547 & -54.11889 & 0.63 & 0.54 & 0.92 \\
        Sextans & 90 & 10.21747 & -1.614722 & 0.63 & 0.54 & 0.88 \\
        Ursa Major I & 97.3 & 10.58133 & 51.92 & 0.58 & 0.5 & 0.83 \\
        Carina & 100 & 6.69353 & -50.96611 & 0.58 & 0.5 & 0.83 \\
        Aquarius II & 107.9 & 22.5654 & -9.328 & 0.5 & 0.42 & 0.79 \\
        Grus I & 120 & 22.9451 & -50.1633 & 0.46 & 0.29 & 0.75 \\
        Fornax & 140 & 2.6665 & -34.4492 & 0.42 & 0.13 & 0.54 \\
        Hercules & 150 & 16.5172 & 12.7917 & 0.38 & 0.13 & 0.54 \\
        Canes Venatici II & 160 & 12.9527 & 34.3200 & 0.33 & 0.08 & 0.46 \\
        Leo IV & 160 & 11.5492 & -0.5333 & 0.33 & 0.04 & 0.46 \\
        Pisces II & 180 & 22.9753 & 5.9525 & 0.21 & 0.04 & 0.42 \\
        Leo V & 180 & 11.5193 & 2.2200 & 0.25 & 0 & 0.375 \\
        Pegasus III & 210 & 22.4063 & 5.4200 & 0.17 & 0 & 0.17 \\
        Canes Venatici I & 220 & 13.4676 & 33.5559 & 0.13 & 0 & 0.13 \\
        Leo II & 250 & 11.2245 & 22.1528 & 0 & 0 & 0 \\
        Leo I & 260 & 10.1411 & 12.3065 & 0 & 0 & 0 \\
        Leo T & 420 & 9.5815 & 17.0514 & 0 & 0 & 0 \\
        Phoenix V & 520 & 0.9916 & 32.3767 & 0 & 0 & 0 \\
        Pisces V & 520 & 0.9916 & 32.3767 & 0 & 0 & 0 \\
        \hline
    \end{tabular}
    \caption{Fraction of the coverage larger than 0.8 for arrival times of the GW spanning over a 24-hour period. Sources are located at the center of 40 nearby galaxies to the Earth and the data are obtained using the \texttt{s20} waveform as an illustrative example.}
    \label{tab:nearby_galaxies}
\end{table*}

To complete the observational prospects with third-generation detectors, we have simulated the GW emission of CCSN explosions occurring at the center of the 40 nearest dwarf galaxies extracted from the Simbad catalogue ~\cite{Wenger:2000sw}. We compare the average coverage for arrival times of the GWs spanning over a 24-hour period obtained in different detector configurations: ET alone, the two CE antennae (CE20-CE40) and the complete network (ET-CE20-CE40). Table \ref{tab:nearby_galaxies} shows, for each nearby galaxy and each network configuration, the fraction of coverage larger than 0.8 obtained with the CCSN waveform \texttt{s20}, chosen as an illustrative case. For example, if a CCSN occurs in the Large Magellanic Cloud there is a 25\% chance that the pipeline will not provide an accurate reconstruction of the ratio $M_{\rm PNS}/R_{\rm PNS}^2$ using the two CE interferometers and ET separately. The pipeline will give precise results 100\% of the time with the combined network ET-CE20-CE40.\\
While the CE20-CE40 network yields only slightly better results than ET when used independently ($\sim$ 1\% improvement only on the 20 nearest galaxies), there is a mean improvement of around 20\% when all third-generation detectors are observing together. If we consider the distance at which the mean coverage is larger than 0.8, then the full network ET-CE20-CE40. allows for the estimation of the PNS parameters at a distance twice as large than considering ET or the CE20-CE40 network alone.

\section{Conclusions}
\label{sec:conclusions}

The complexity of the physics involved in the explosion and the opacity of its environment at the onset of the collapse make the fate of massive stars rather difficult to study with electromagnetic observations. The GW signal emitted in those events, although weak, provides unperturbed information about the dynamics of the explosion. If a Galactic CCSN happens when the current generation of GW detectors are acquiring data, one should be able to reconstruct a large fraction of the highly stochastic GW signal, estimate the total GW energy emitted and measure the GW signal spectrum. The next step is to infer some properties of the progenitor and of the compact remnant, as proposed in~\cite{Torres:2019b,Sotani:2021ygu}, which demonstrate that the time evolution of the buoyancy-driven g-mode excitations are linked to the physical properties of the compact remnant through universal relations. Building on the proof-of-principle study we initiated in~\cite{Bizouard:2020sws}, we have presented in this paper a realistic CCSN inference pipeline that allows to analyze coherently the data of a multi-detector network to extract PNS physical parameters from the collected GW data. More precisely, we use a standard likelihood function to combine the data from two or more GW detectors and build time-frequency maps in which the PNS oscillation modes are tracked with a LASSO regression algorithm. Focusing on the main g-mode, we have shown that the algorithm is able to measure the time evolution of $M_{\rm PNS}/R_{\rm PNS}^2$ and its 95\% confidence interval. To quantify the quality of the estimation we use the fraction of true values that are within the 95\% confidence interval. This index allows us to evaluate the maximal distance at which we could reconstruct physical parameters of the PNS in different GW detector network configurations. This depends on the arrival time of the signal on Earth, or on how large is the detector's response to the signal. We have shown that, on average, the current generation of detectors at their design sensitivities are capable of measuring $M_{\rm PNS}/R_{\rm PNS}^2$ of a Galactic CCSN for most of the progenitors considered in the study. We have also studied the gain of having a large network of second-generation detectors on Earth, showing that the five detector network (three LIGOs, Virgo and KAGRA) consistently improves the performance of our pipeline with respect to smaller combinations of detectors. We point out that the different duty cycles of each of the second generation GW interferometers do not always allow a coherent analysis with the complete network. For example, during the O3b run both LIGO interferometers and Virgo operated simultaneously for ~51\% of the time \citep{Abbott:2021djp}. A larger number of detectors would naturally increase the coincident operation times and thus further improve the observational prospects. Finally, we have repeated the analysis for a network composed of the planned third-generation detectors, Einstein Telescope and Cosmic Explorer. ET in its triangular configuration is composed of three co-located detectors, allowing to estimate by itself the physical parameters for a CCSN source several hundreds of kpc from Earth. Our results also indicate that a network with the two CE detectors yields a similar performance and that a global network of the three third-generation detectors (CE20-CE40-ET) would typically double the distance reach. 

We end by pointing out that devising a more advanced tracking algorithm than the one used in this paper is desirable. This should help to better reconstruct the oscillation frequency spectrum of the PNS when modes other than the $^2g_2$ mode are also carrying significant amounts of GW energy. The spectrograms constructed from CCSN numerical simulations show that this might be particularly important for tracking a downward secondary feature associated to the $^2g_3$ mode and its avoided crossing with the dominant $^2g_2$ mode. In addition, reconstructing the signature of the SASI, at lower frequencies, is also potentially interesting and could provide information about the time evolution of the shock radius and about the total mass inside the shock using universal relations.

\bigskip\noindent\textit{Acknowledgments.--} Work supported by the Spanish Agencia Estatal de Investigaci\'on (Grants No. PGC2018-095984-B-I00 and PID2021-125485NB-C21) funded by MCIN/AEI/10.13039/501100011033 and ERDF A way of making Europe, by MCIN and Generalitat Valenciana with funding from European Union NextGenerationEU (PRTR-C17.I1, Grants ASFAE/2022/003 and ASFAE/2022/026), by the Generalitat Valenciana (PROMETEO/2019/071), and by the European Union’s Horizon 2020 research and innovation (RISE) programme (H2020-MSCA-RISE-2017 GrantNo.~FunFiCO-777740).
N.C. acknowledges support from National Science Foundation grant PHY-1806990.
R.M.\ and P.M.-R.\ gratefully acknowledge support by the Marsden grant MFP-UOA2131 from New Zealand Government funding, administered by the Royal Society Te Ap\={a}rangi, funding from DFG Grant KI 1443/3-2 and thank the Centre for eResearch at the University of Auckland for their technical support.
P.C.-D.\ and M.O.\ acknowledge the support of the Spanish {\it Ramon y Cajal} programme (RYC-2015-19074 and RYC-2018-024938-I, respectively) supporting their research.

\bibliographystyle{apsrev4-1}
\bibliography{biblio}

\end{document}